\documentclass[a4paper,11pt]{article}

\pdfoutput=1

\usepackage{jinstpub}
\usepackage{comment}
\usepackage{siunitx}
\usepackage{graphicx}
\usepackage{color}
\usepackage{subcaption}
\usepackage{soulutf8}
\usepackage[T1]{fontenc}

\usepackage{lineno}

\usepackage{textgreek} 
\usepackage{tabulary}
\usepackage{wasysym} 
\usepackage{setspace}
\usepackage{latexsym} 
\usepackage[export]{adjustbox} 
\usepackage{makecell} 
\usepackage{xfrac} 
\usepackage{float} 

\newcommand*{\affmark}[1][*]{\textsuperscript{#1}}

\title{Characterizing Novel Indium Phosphide Pad Detectors with Focused X-ray Beams and Laboratory Tests}

\author{E.R.~Almazan{\normalfont\it\affmark[a,]}
{\normalfont\affmark[1]},\note{Corresponding author}}
\author{A.~Affolder{\normalfont\it\affmark[a]},}
\author{I.~Dyckes{\normalfont\it\affmark[b]},}
\author{V.~Fadeyev{\normalfont\it\affmark[a]},}
\author{M.~Hance{\normalfont\it\affmark[a]},}
\author{M.~Jadhav{\normalfont\it\affmark[c]},}
\author{S.~Kim{\normalfont\it\affmark[c,d]},}
\author{T.~McCoy{\normalfont\it\affmark[a]},}
\author{J.~Metcalfe{\normalfont\it\affmark[c]},}
\author{J.~Nielsen{\normalfont\it\affmark[a]},}
\author{J.~Ott{\normalfont\it\affmark[a]},}
\author{L.~Poley{\normalfont\it\affmark[e]},}
\author{T.(K.-W.)~Shin{\normalfont\it\affmark[a]},}
\author{D.~Sperlich{\normalfont\it\affmark[f]},}
\author{A.~Sumant{\normalfont\it\affmark[c]}}

\affiliation[a]{Santa Cruz Institute for Particle Physics, University of California at Santa Cruz, 1156 High Street, Santa Cruz, CA, USA}
\affiliation[b]{Lawrence Berkeley National Laboratory,
1 Cyclotron Rd, Berkeley, CA, USA}
\affiliation[c]{Argonne National Laboratory, Lemont, IL, US}
\affiliation[d]{Department of Chemical Engineering, University of Illinois at Chicago, USA}
\affiliation[e]{TRIUMF,4004 Wesbrook Mall, Vancouver, B.C., Canada}
\affiliation[f]{Physikalisches Institut, Albert-Ludwigs-Universitat Freiburg, 12
13 Hermann-Herder-Strasse, Freiburg im Breisgau, Germany}

\emailAdd{etalmaza@ucsc.edu}

\abstract{
Future tracking systems in High Energy Physics experiments will require large instrumented areas with low radiation length.  Crystalline silicon sensors have been used in tracking systems for decades, but are difficult to manufacture and costly to produce for large areas.  We are exploring alternative sensor materials that are amenable to fast fabrication techniques used for thin film devices. Indium Phosphide pad sensors were fabricated at Argonne National Lab using commercially available InP:Fe 2-inch mono-crystal substrates. Current-voltage and capacitance-voltage characterizations were performed to study the basic operating characteristics of a group of sensors.  Micro-focused X-ray beams at Canadian Light Source and Diamond Light Source were used to study the response to ionizing radiation, and characterize the uniformity of the response for several devices. Electrical test results showed a high degree of performance uniformity between the 48 tested devices. X-ray test beam results showed good performance uniformity within tested devices after accounting for spatially-local defects and edge fields. This motivates further studies into thin film devices for future tracking detectors.
}

\keywords{
 Indium Phosphide; radiation detectors; x-ray beam; HEP; tracking}

\begin{document}
\maketitle
\flushbottom

\section{Introduction}

Experimental High Energy Physics progress in recent decades has relied on solid-state ionization detection technologies for precision charged-particle tracking.  Advancements in micro-fabrication and semiconductor properties led to the current generation of large silicon trackers currently in use at the Large Hadron Collider and other facilities.  However, these systems take many years of development and several years of construction time due to the complexity and production throughput limitations for unique components.

Alternative sensor designs based on thin-film fabrication technology can help to alleviate some of these bottlenecks. A survey of possible materials and fabrication techniques was performed in Ref. \cite{Kim_2023}, in which several promising materials were identified for further study. Indium Phosphide (InP) is one such material.  It is amenable to the thin-film growth techniques, and it features a fairly high electron mobility of 4600 $\mathrm{cm^2/(V\cdot s)}$ \cite{Kim_2023}, more than a factor of three higher than that of silicon, lending itself to fast charge collection and accurate timing measurements.

As a first step of the exploration, we fabricated pad devices on 2-inch wafers using single-crystal commercially available wafers. The substrate thickness was around 350 $\mu m$. The presence of iron dopant serves as a trapping defect for hole movement. The single-crystal nature of these devices implies having fewer defects than the amount we may get with thin-film growth techniques. However, the relatively large thickness leads to smaller electric fields and potentially more trapping. We considered these devices to be a good starting point for the basic characterization of the material response to the ionizing radiation and high-voltage operations expected in an ionization-sensitive device.

By probing monocrystalline InP uniformity, we aim to establish a baseline against which to compare eventual thin-film InP sensors. For inter-device uniformity, we performed basic electrical tests (IV and CV) across devices fabricated from the same wafer. A probe of intra-device uniformity was done with a focused x-ray beam, measuring the photoresponse as a function of beam position on the device, and correlating the photocurrent to geometrical features of the device.

\section{Sample Fabrication and Basic Properties}
\label{sectionBasic}

Detector fabrication was carried out in the cleanroom facilities of The Center for Nanoscale Materials at Argonne National Laboratory. Initial wafers were polished on one side and etched on another. There were two layers of metallization deposited, on the back side and the front side of the wafers. The metal coverage on the back side was continuous. The front side included patterning to make a two-dimensional array of diodes on 5 mm pitch. The metal thickness was relatively small, of 110 nm. The central pad was 2x2 $\mathrm{mm^2}$ in size. The guard ring was 100 $\mu$m wide, spaced from the central pad by a gap of 100 $\mu$m. To prevent electrical breakdown from conductor geometry, the central pad and guard rings were designed with curved edges. The guard ring width and distance to the central pad was kept constant at all locations. The radius of the curved pad corner was 100 $\mu$m (Figure~\ref{fig:topside_metrology1}). Further fabrication details are given elsewhere \cite{Kim_2024}.

Some of the devices had an additional metal openings in the center, of 150 $\mu$m diameter to facilitate studies with laser (Figure~\ref{fig:topside_metrology2}). Additionally, multi-pad devices were included, fabricated from a separate identical wafer. They were tested separately, and will be presented in a different publication.

While the composition of material is InP:Fe, devices and their bulk material will be referred to as InP for the sake of simplicity.

\section{Geometry Measurements}
\label{sectionGeom}

\begin{figure}
\begin{subfigure}{.5\textwidth}
    \centering
    \includegraphics[width=.95\linewidth]{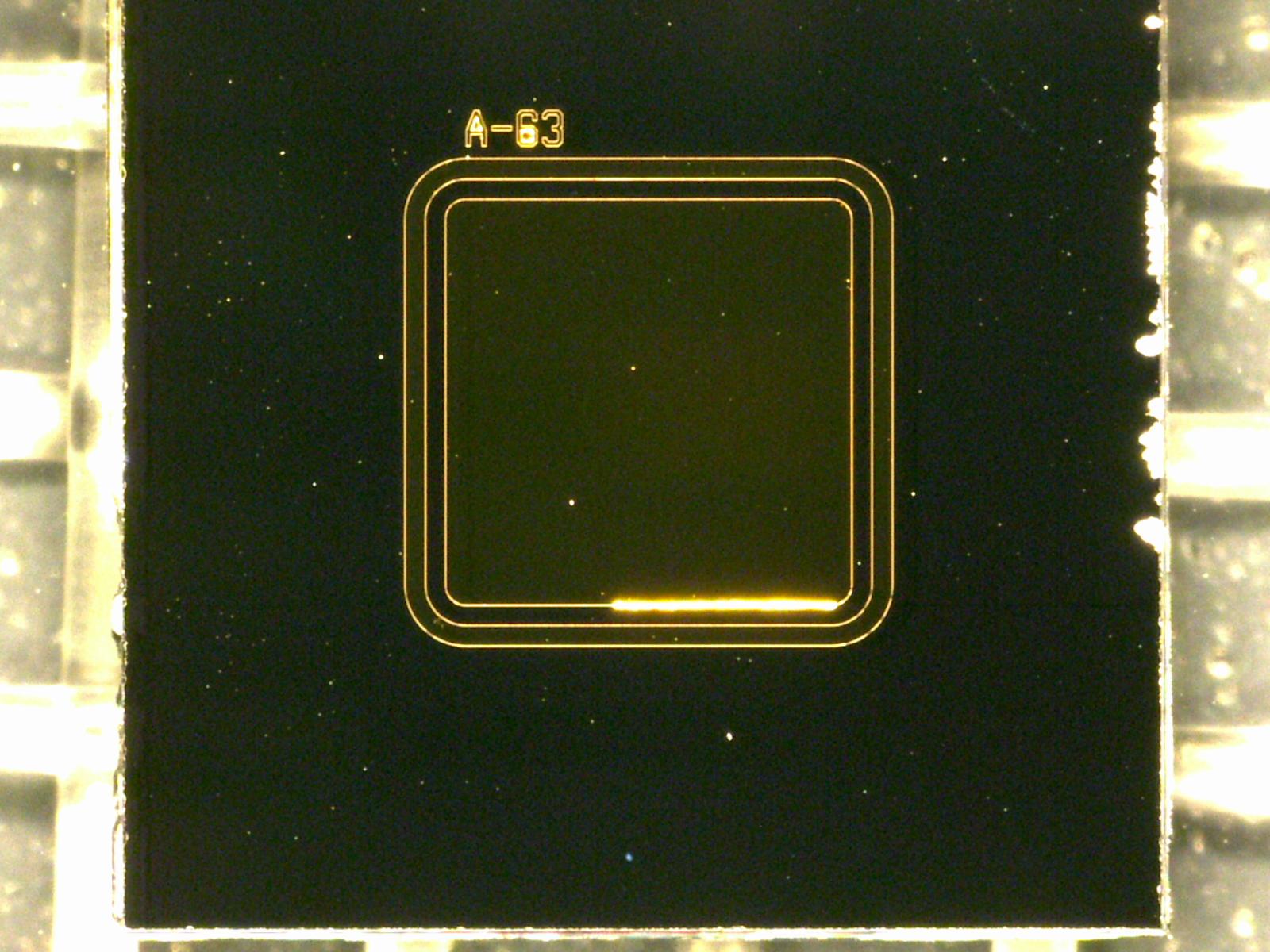}
    \caption{Photograph of A63\_NH}
    \label{fig:topside_metrology1}
\end{subfigure}
\begin{subfigure}{.5\textwidth}
    \centering
    \includegraphics[width=.95\linewidth]{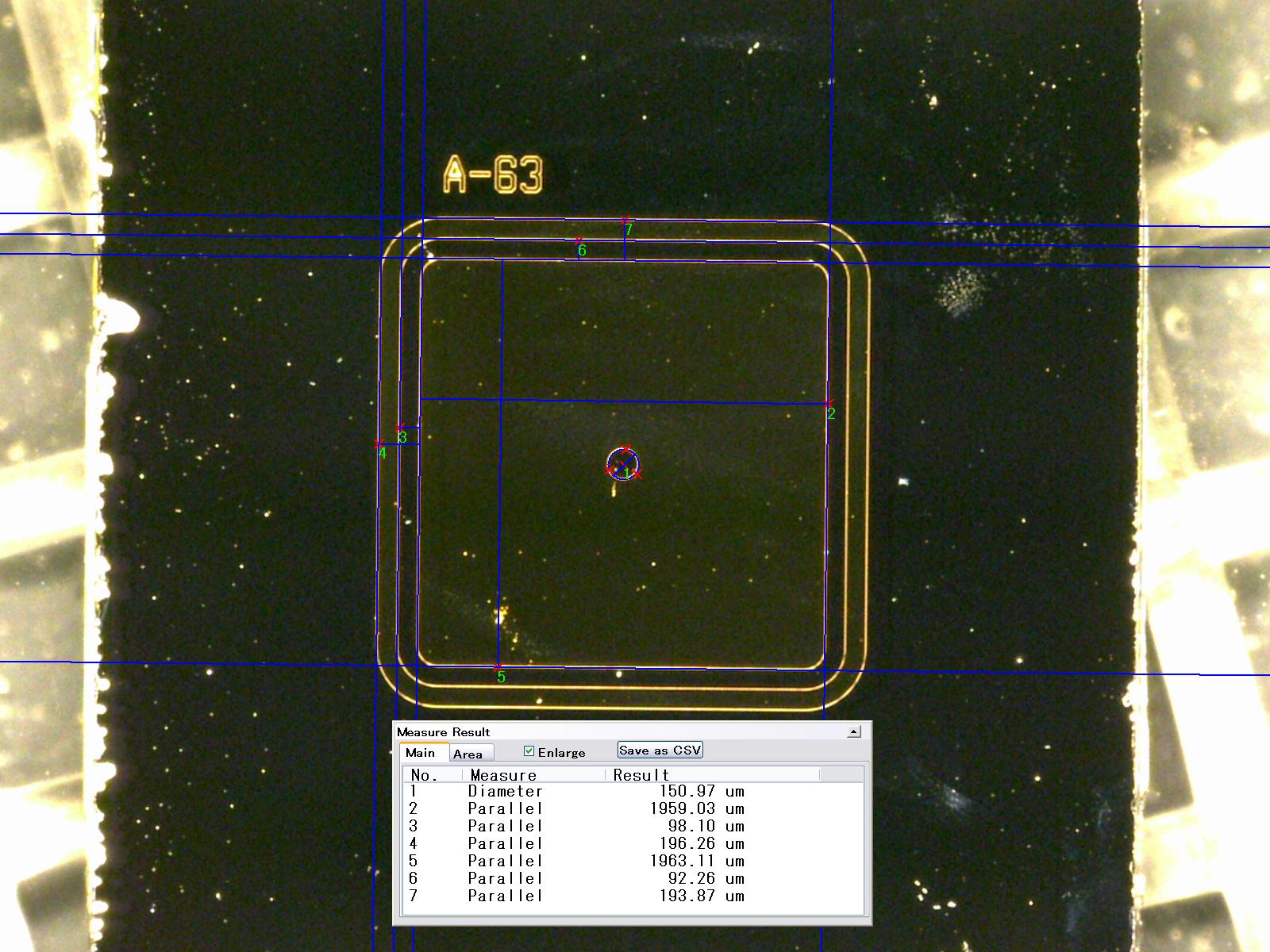}
    \caption{Photograph of A63\_H with metrological data}
    \label{fig:topside_metrology2}
\end{subfigure}
\caption{Frontside profile photos of typical devices, one with device solid central pad metallization (left) and another containing a central hole in the pad (right)}
\label{fig:topside_metrology_total}
\end{figure}

\begin{figure}
    \centering
    \includegraphics[width=0.8\textwidth]{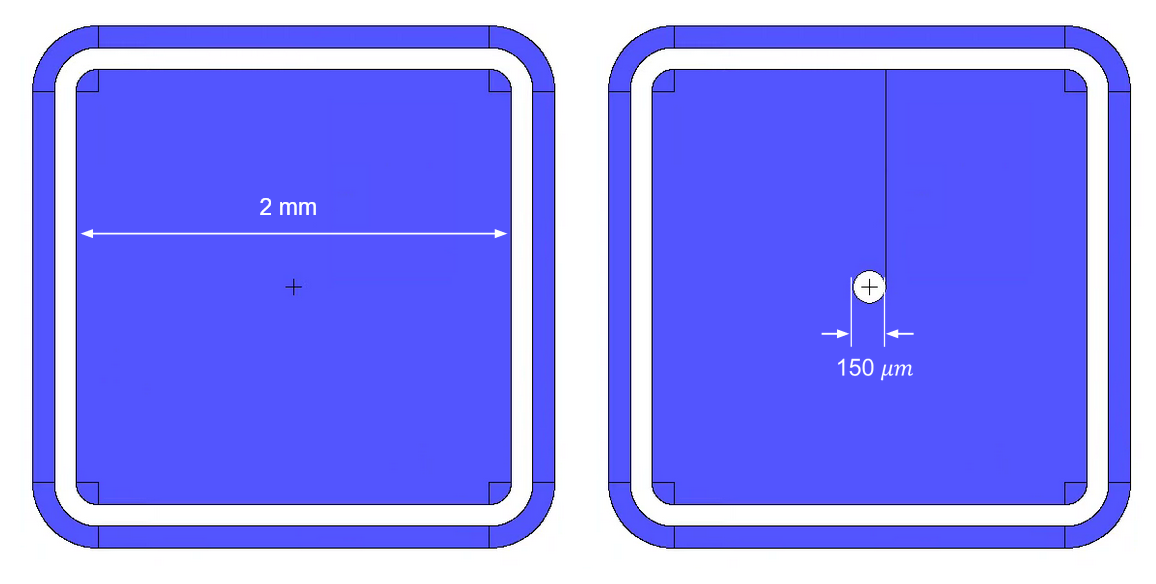}
    \caption{Shadow masks used to pattern the frontside conductors of single-pad devices.\\
    The left image shows the "no-hole" (NH) conductor pattern, while the right image shows the "hole" (H) conductor pattern. Both patterns contain a central pad and guard ring, save for the 150 $\mu$m diameter hole in the H central pad.}
    \label{fig:diagram_top}
\end{figure}

We first performed metrological assessment of the device geometry. Frontside photos of all devices were taken using the Keyence VHX-500 optical inspection station. Central pad and guard ring dimensions were measured from these photos (Figure~\ref{fig:topside_metrology_total}). These measurements are consistent with the layout target dimensions within the expected accuracy of these measurements, of a few percent (Figure \ref{fig:diagram_top}).

\begin{figure}
\begin{subfigure}{.5\textwidth}
    \centering
    \includegraphics[width=.95\linewidth]{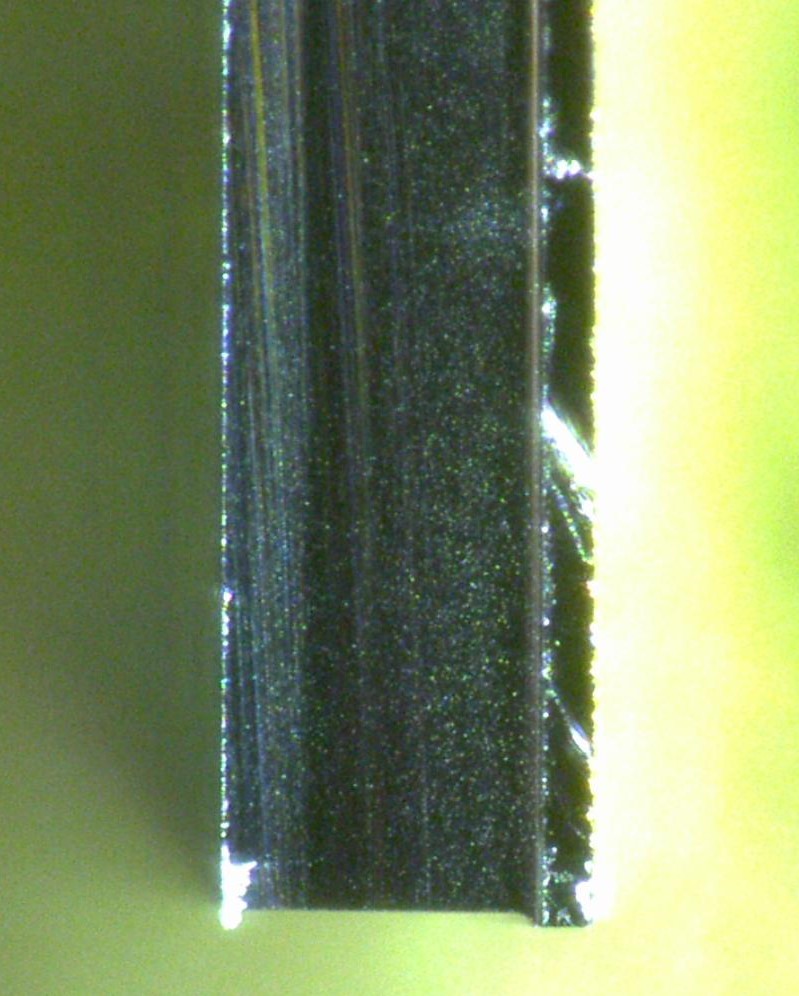}
    \caption{An edge profile photograph of A51\_NH}
    \label{fig:edge_metrology1}
\end{subfigure}
\begin{subfigure}{.5\textwidth}
    \centering
    \includegraphics[width=.95\linewidth]{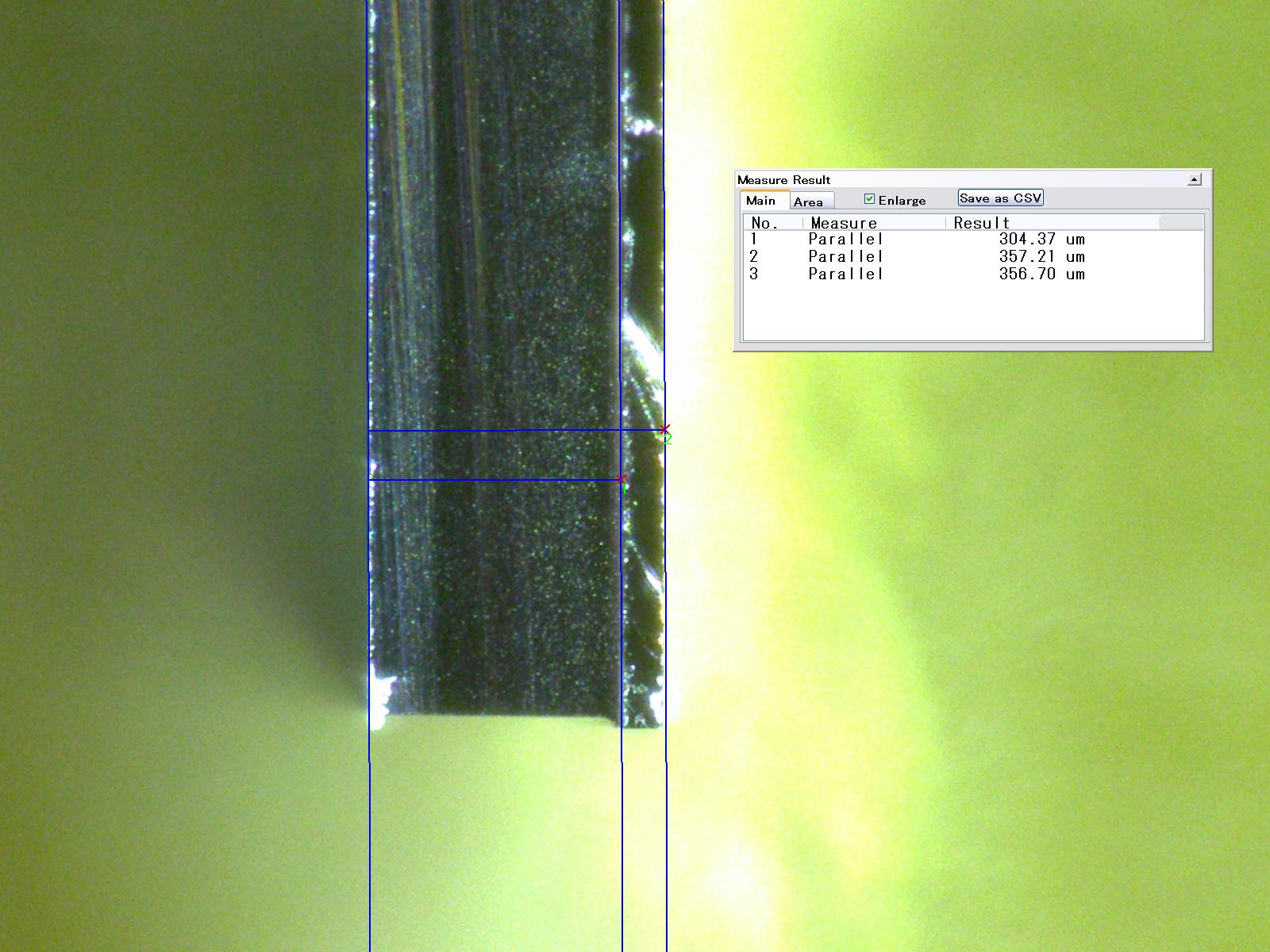}
    \caption{An edge profile photograph of A51\_NH with metrological data}
    \label{fig:edge_metrology2}
\end{subfigure}
\caption{Edge profile photos of a typical device}
\label{fig:edge_metrology_total}
\end{figure}

An edge-profile photo was taken of device A51\_NH to verify device thickness (Figure~\ref{fig:edge_metrology_total}). The thickness of the InP wafer, based on the Quality Test Report provided by the vendor, PAM-XIAMEN, is expected to be between 325 $\mu$m to 375 $\mu$m. Edge profile metrological data (Figure \ref{fig:edge_metrology2}) corroborates with the Quality Test Report with a measured InP thickness of 357 $\mu$m.

\begin{figure}
    \centering
    \includegraphics[width=0.8\textwidth]{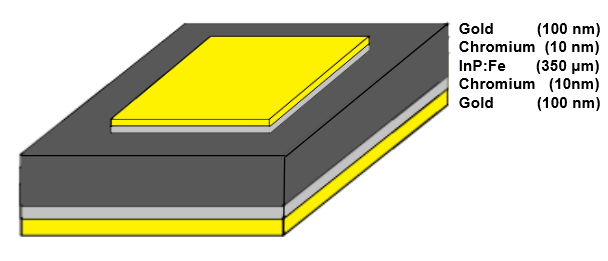}
    \caption{A stackup diagram of InP devices architecture.}
    \label{fig:diagram_edge}
\end{figure}

\section{Electrical Tests}
\label{sectionIVCV}

The devices were first characterized with measurements of their leakage current and capacitance. These parameters were derived from standard current-voltage (IV) and capacitance-voltage (CV) scans.

The following measurements were performed in a manual probe station. 

For IV scans, a Keithley 2657A SourceMeter unit was used to supply the bias voltage through the probe station chuck to the backplane of the device and simultaneously to measure the total current. The device was connected to ground by probe needle from the central pad of the device.

For CV characterization, the device was connected by probe needle from the central pad of the device through a current-potential decoupling box to an Agilent E4980A Precision LCR meter, with the Keithley 2657A supplying a DC offset through the probe station chuck. The capacitance was recorded at AC test frequencies of 1 kHz, 10 kHz, 100 kHz, and 1 MHz.

\subsection{IV Measurements}
\label{subsectionIVmeas}

\begin{figure}
\begin{subfigure}{.5\textwidth}
    \centering
    \includegraphics[width=.95\linewidth]{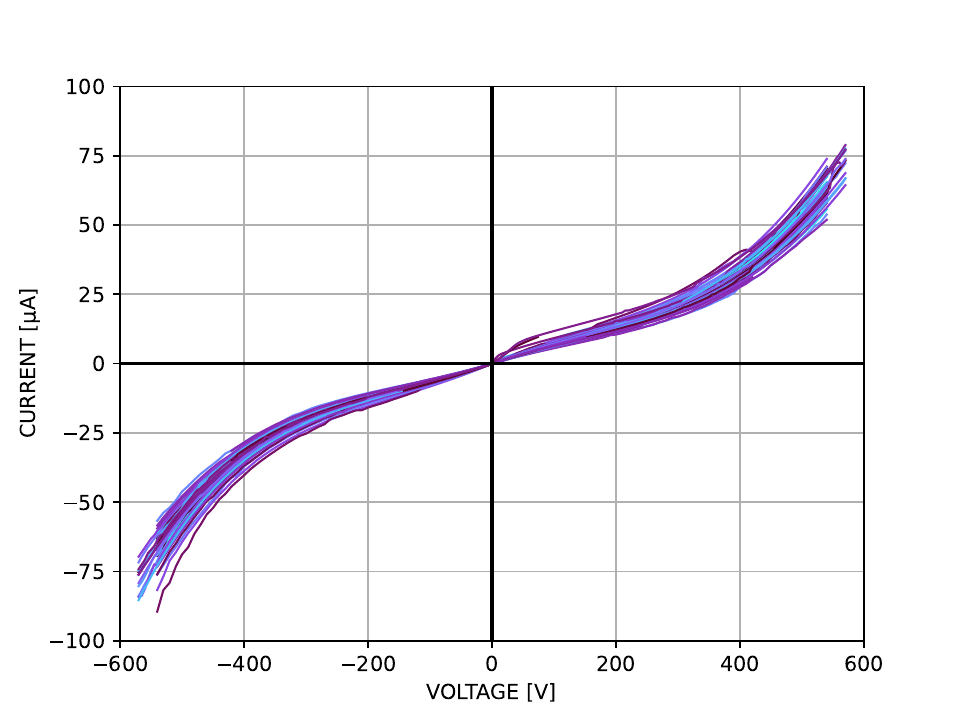}
    \caption{Leakage Current vs Voltage}
    \label{fig:iv_floating_plot}
\end{subfigure}
\begin{subfigure}{.5\textwidth}
    \centering
    \includegraphics[width=.95\linewidth]{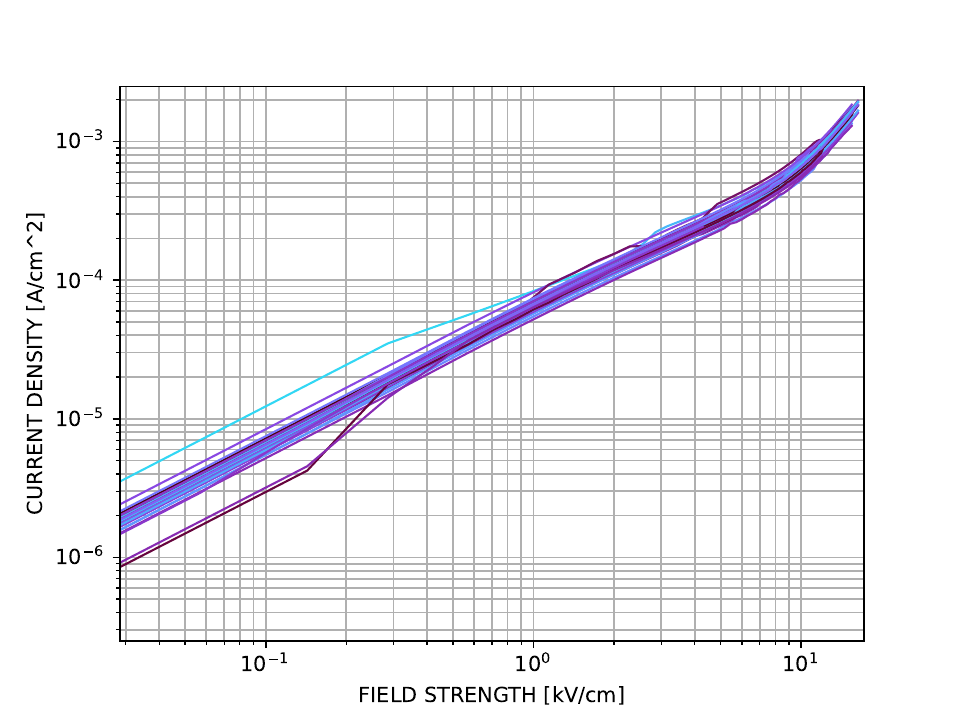}
    \caption{Leakage Current Density vs Field Strength} 
    \label{fig:iv_floating_scaled_plot}
\end{subfigure}
\caption{IV curves of the full set of 48 InP devices.  Each device corresponds to a distinct colored line. The majority of IV curves span from -600 V to 600 V. All IV curves above were recorded with a floating guard ring.  In the second figure, IV measurements are shown from 1-600 V with logarithmic horizontal and vertical axes, with current density and field strength replacing current and voltage respectively. The current density and field strength are calculated assuming the architecture of an ideal geometry, using the central pad area for the current normalization.}
\label{fig:iv_floating_total}
\end{figure}

The InP devices can be thought of as capacitors with uniform dielectric bulk. The uniformity of the bulk implies the lack of a space-charge region that widens or narrows in response to bias polarity. Measured IV curves are therefore expected to be symmetric with respect to a sign flip of the voltage. The overall behavior of IV measurements shown in Figure \ref{fig:iv_floating_plot} is consistent with this representation. No breakdown voltage is found between -600 V and 600 V. 

The IV behavior is linear until around 10 kV/cm, which corresponds to +350 V (Figures \ref{fig:iv_floating_plot} and \ref{fig:iv_floating_scaled_plot}). This voltage, corresponding to the transition from ohmic to non-ohmic behavior, is chosen as the operating voltage for summarized properties of these devices.

The power dissipation of InP devices at 350 V is estimated to be 9.08 mW, calculated from the measured leakage current with only the central pad grounded (Table \ref{IV_table}). Assuming the majority of current flows through only the volume of the device under the central pad, the active volume can be approximated as .0014 $cm^3$ using information from Section \ref{sectionGeom}. InP then has an approximate power density of 6.48 W cm$^{-3}$.

\subsection{Breakdown Voltage}
\label{subsectionBreakdown}

The IV test range described in previous sections was limited to $\pm$ 600 V.  The breakdown voltage for these devices is estimated by finding the largest voltage the pads are capable of receiving before the resulting leakage current rapidly increases between 300 V and 600 V. For these breakdown voltage tests, the aforementioned -600 V and +600 V bounds were instead stretched to $\pm$1050 V, using the same instrumentation and equipment setup as the other IV floating tests (see Section \ref{sectionIVCV}).

Breakdown in some given material, such as silicon, may vary between devices depending on fabrication and design artifacts. For example, fields can build up on sharp edges in the structures of the electrode or implant doping variation. As a base metric, however, the expected breakdown of pure bulk Si is approximately 3$\cdot$10$^5$V/cm \cite{Semiconductor}.

\begin{figure}[!ht]
\begin{subfigure}{.5\textwidth}
    \centering
    \includegraphics[width=.95\linewidth]{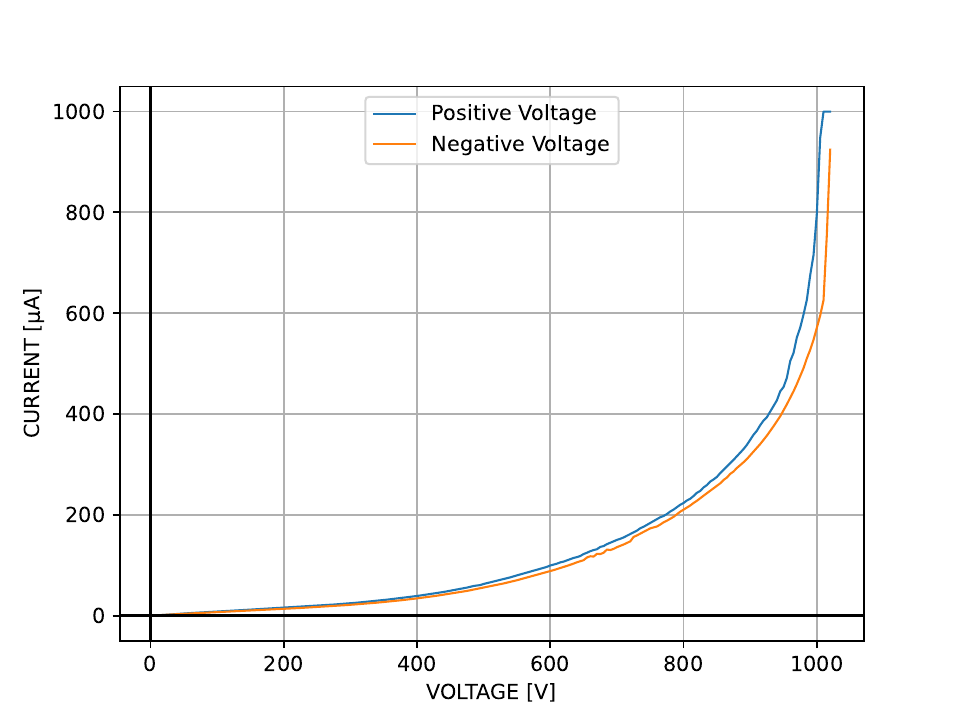}
    \caption{Linear Axes}
    \label{fig:iv_breakdown}
\end{subfigure}
\begin{subfigure}{.5\textwidth}
    \centering
    \includegraphics[width=.95\linewidth]{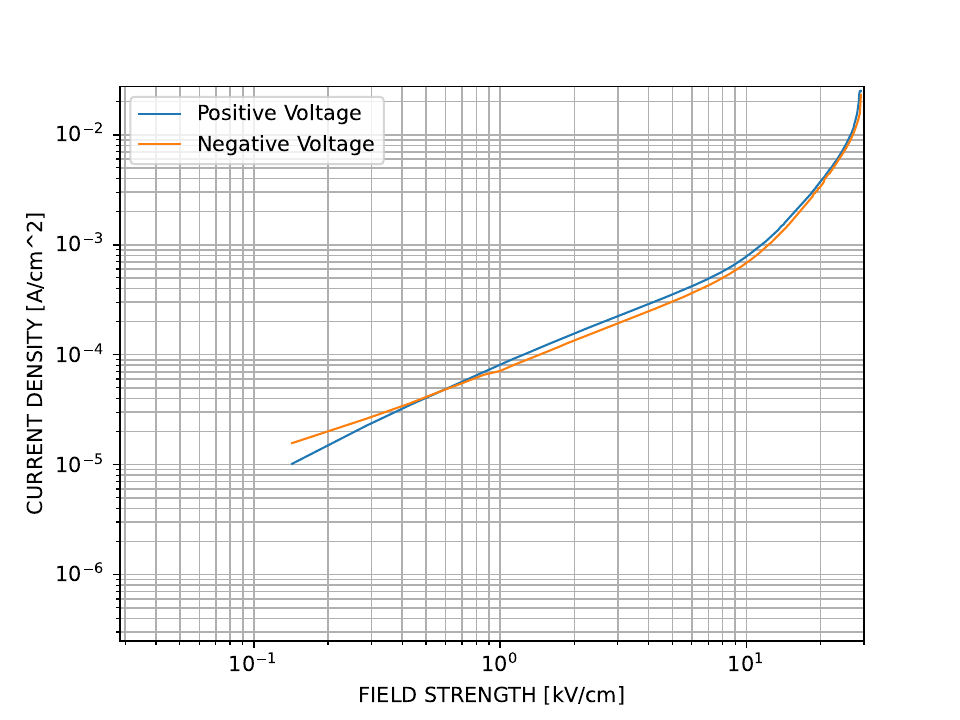}
    \caption{Log Axes} 
    \label{fig:iv_breakdown_scaled}
\end{subfigure}
\caption{IV plots in extended voltage range for one device: linear axes in the left plot and logarithmic axes in the right plot. Both positive and negative applied voltage polarity are shown.
The flattened end of the IV curve comes from hitting compliance on the SourceMeter, and does not indicate any feature of InP.} 
\label{fig:iv_breakdown_total}
\end{figure}

An analysis of InP's breakdown trend is shown in Figure \ref{fig:iv_breakdown_total}, Figure (a) displays current behavior reaching breakdown at about $\pm$ 1010 V, thus breakdown was never approached during standard IV measurements. This breakdown voltage is equivalent to roughly 28.9 kV/cm field strength, assuming an ideal capacitor geometry using the central pad area for current normalization. Figure (b) shows the field strength as a function of current density, showing the expected positive correlation between an increase in the electric field's intensity and the density of the current.

\subsection{IV Device Variability}
\label{subsectionIVuniform}

\begin{figure}
\begin{subfigure}{.5\textwidth}
    \centering
    \includegraphics[width=.8\linewidth]{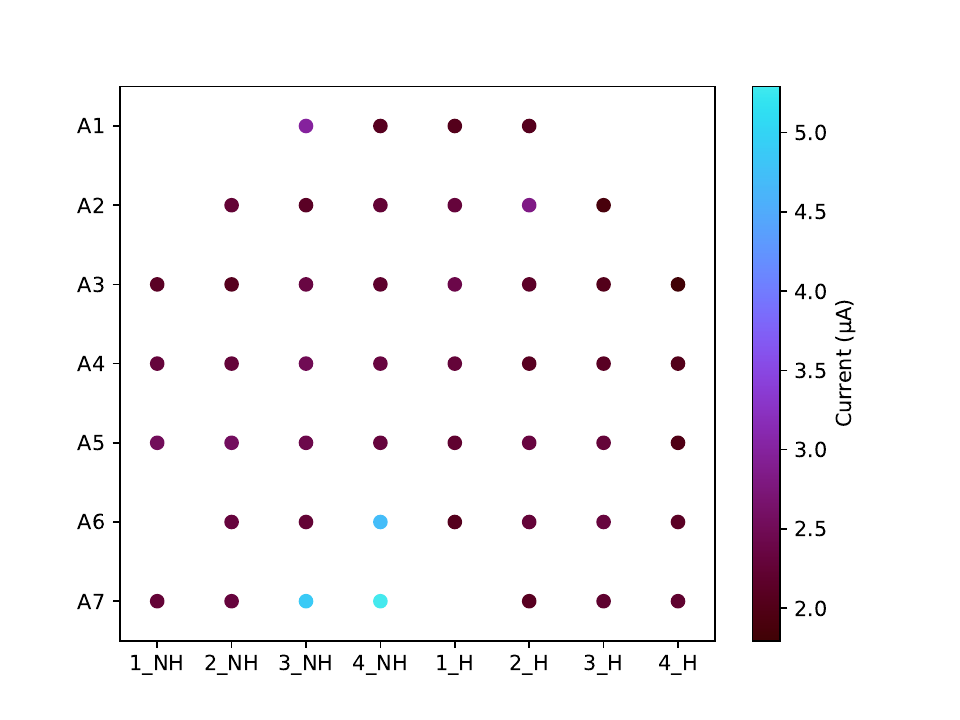}
    \caption{V = +30 V}
    \label{fig:iv_floating_colormap_30}
\end{subfigure}
\begin{subfigure}{.5\textwidth}
    \centering
    \includegraphics[width=.8\linewidth]{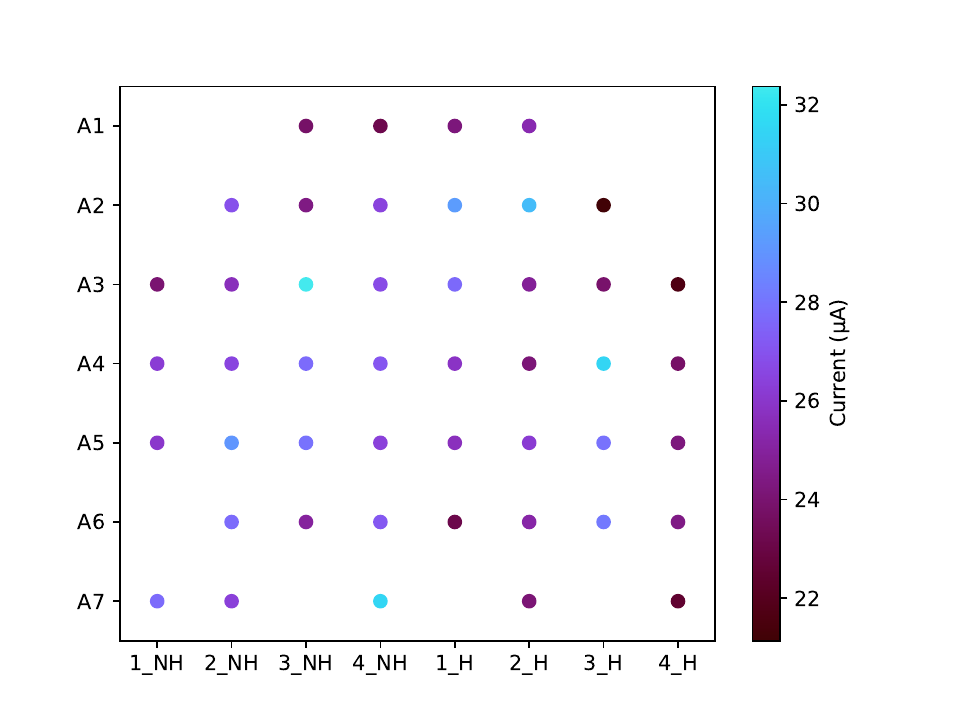}
    \caption{V = +350 V}
    \label{fig:iv_floating_colormap_350}
\end{subfigure}
\caption{Leakage Current Colormaps.\\
Each device/point is colored by their current at each listed bias, with the color axis scaled by the minimum and maximum leakage currents measured across devices at said bias. Their positions in the grid correspond to their positions in the original wafer before dicing and separation. \\
The naming convention for all devices is by their row + column in the the grid. For example, the bottom left device on both scatter plots is labelled as A7 + 1\_NH = A71\_NH. The "high-current" device set is comprised of A64\_NH, A73\_NH, and A74\_NH.}
\label{fig:iv_floating_colormaps}
\end{figure}

Device IV responses are qualitatively uniform, save for three exceptions. Three devices were identified with notably higher current at low positive voltage: A64\_NH, A73\_NH, and A74\_NH (See Figure \ref{fig:iv_floating_colormaps} and the label explanation therein.). These devices will be referred to as the "high-current" devices for the sake of clarity. The high-current devices are identified by a higher leakage current at low positive bias voltage compared to other devices from the same wafer (Figure \ref{fig:iv_floating_colormap_30}). This high-current behavior is absent at higher bias voltages (Figure \ref{fig:iv_floating_colormap_350}). Notably, the high-current devices neighbored each other physically in the original wafer before dicing. We hypothesize then that the cause of high-current device behavior is due to a local physical defect shared by these neighboring devices. While the exact nature of this defect is under investigation, its existence motivates the exclusion of high-current devices from assessing leakage current uniformity across devices. 

\begin{figure}
\begin{subfigure}{.5\textwidth}
    \centering
    \includegraphics[width=.95\linewidth]{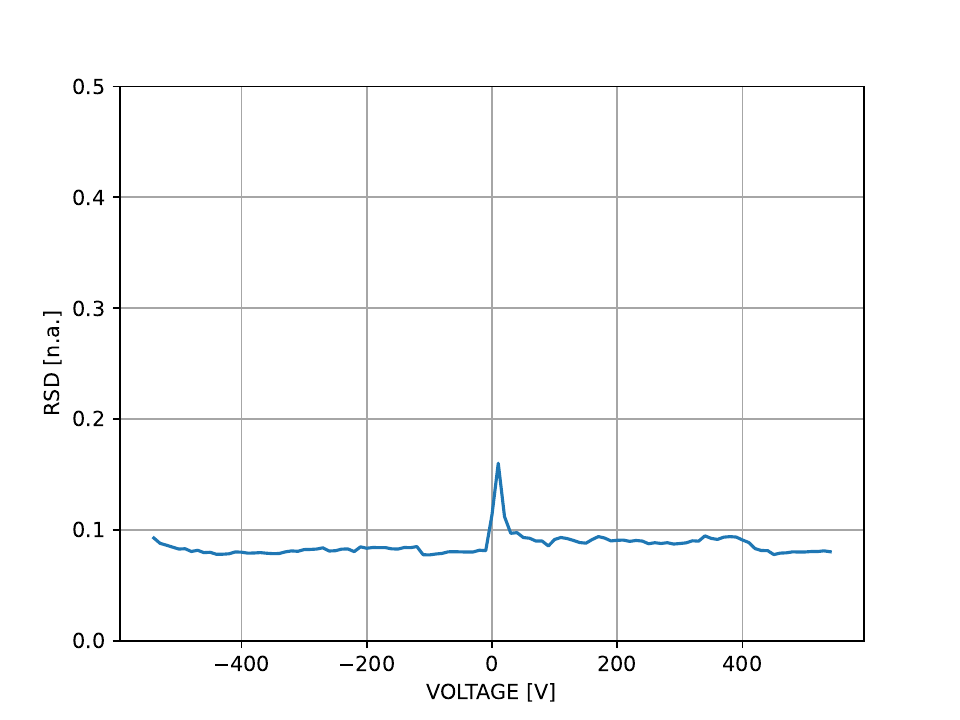}
    \caption{Leakage Current Relative Standard Deviation}
    \label{fig:iv_rsd}
\end{subfigure}
\begin{subfigure}{.5\textwidth}
    \centering
    \includegraphics[width=.95\linewidth]{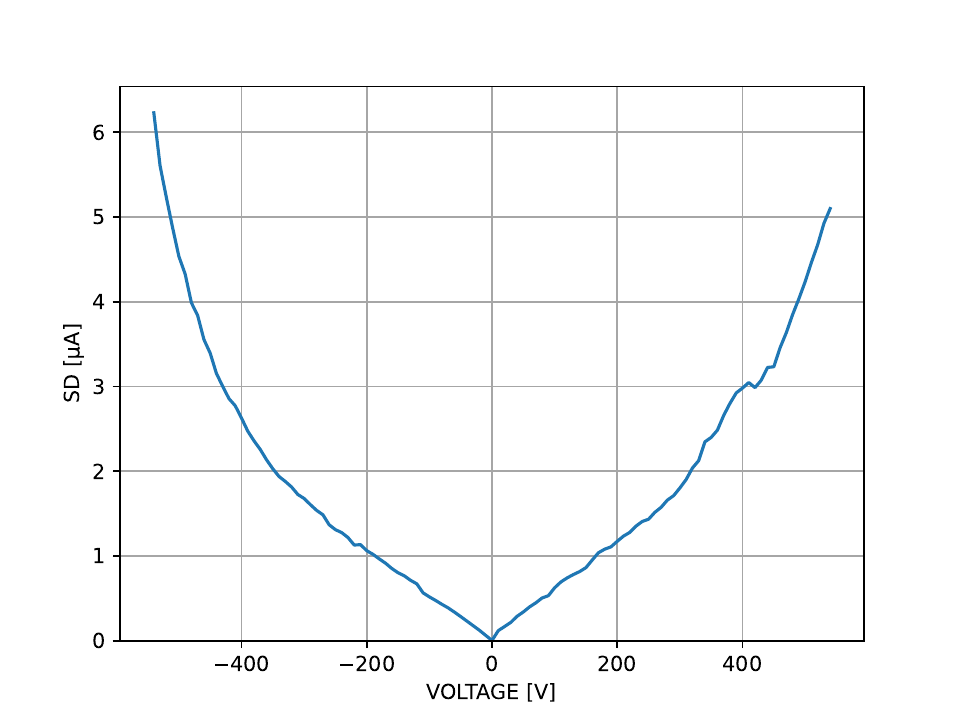}
    \caption{Leakage Current Standard Deviation}
    \label{fig:iv_std}
\end{subfigure}
\caption{Measures of leakage current variance, excluding high-current devices. The RSD spike close to 0V in the first image can be attributed to the measurement error being relatively large at low value of leakage current. The average RSD across all voltages is 0.085.}
\label{fig:iv_cov_std_total}
\end{figure}

As InP is a novel material for charged particle detection, quantitative measures of device uniformity are essential in assessing the consistency of future InP detector responses. Relative Standard Deviation (RSD), defined as the standard deviation (SD) divided by the population mean, is a standardized measure of dispersion in a sample population commonly used in quality assurance studies. An advantage to using leakage current RSD as a measure of leakage current uniformity over SD is that it is dimensionless. RSD then is independent of the leakage current mean and sample population size at each voltage, allowing for the comparison of leakage current uniformity across the full range of measured bias voltages. Qualitatively, RSD lacks a dependence on bias magnitude or polarity, save for the exception at low positive bias voltage due to high relative noise compared to low leakage current (Figure \ref{fig:iv_rsd}). The near constant RSD implies that the leakage current SD is dependent on the leakage current magnitude and not on the test system. 

RSD encapsulates the quadratic sum of all sources of variance, from random errors to device non-uniformity. For example, the absence of climate control in the probe station leads to variations in leakage current due to temperature variations in the lab, taken over the course of a year. For this reason, the 0.092 RSD measured should be taken as an upper limit for device-to-device variation.

\subsection{CV Measurements}
\label{subsectionCVmeas}

\begin{figure}
\begin{subfigure}{.5\textwidth}
    \centering
    \includegraphics[width=.95\linewidth]{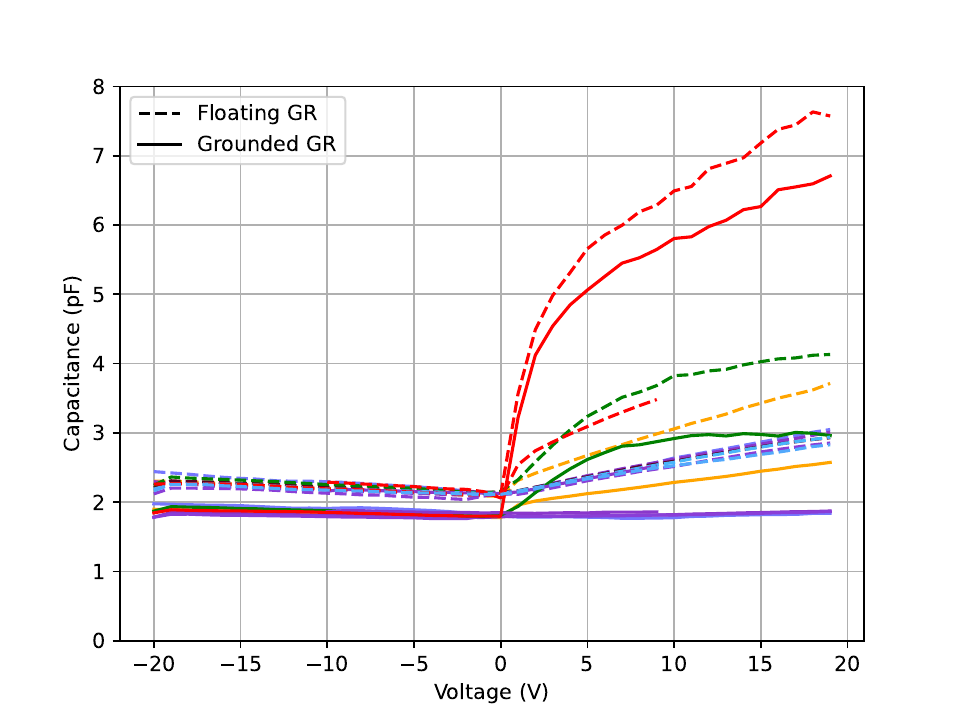}
    \caption{CV measured at f = 1 kHz}
    \label{fig:cv_1khz}
\end{subfigure}
\begin{subfigure}{.5\textwidth}
    \centering
    \includegraphics[width=.95\linewidth]{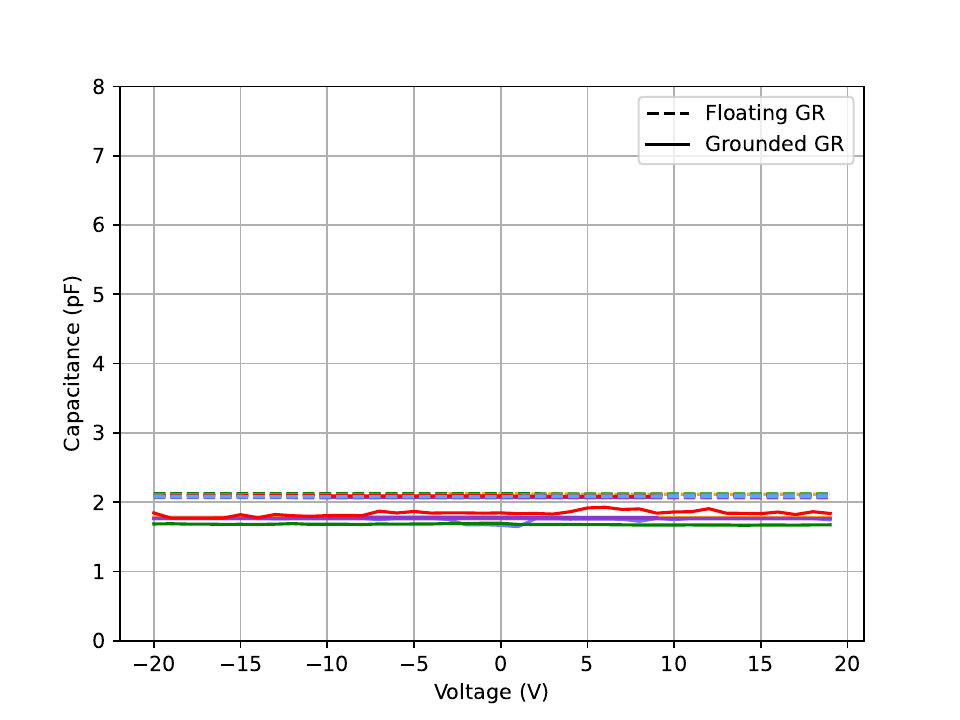}
    \caption{CV measured at f = 1 MHz.}
    \label{fig:cv_1mhz}
\end{subfigure}
\caption{CV Measurements at 1 kHz and 1 MHz. Dashed and solid lines correspond with CV measurements taken with a floating and grounded guard ring correspondingly. The voltage dependence of the capacitance decreases at large frequencies.
Nominal CV measurements are colored from light-blue to purple according to a one-to-one mapping from device position to the gem colormap from the scientific colormap python package CMasher. High-current measurements are colored red and exhibit higher capacitances at 1 kHz. Measurements for A63\_NH and A51\_NH are colored green and yellow respectively to show their aberrant behavior.}
\label{fig:cv_total}
\end{figure}

Device capacitance exhibits voltage dependence at low AC frequencies, which is lost as the frequency is raised (Figure \ref{fig:cv_1mhz}). At low frequencies, the high-current devices had a noticeably larger capacitance than other devices, with floating GR capacitances above 3 pF at +5V. 

A simple model of the devices is a capacitor with parallel plates and dielectric in-between. For
this model, the capacitor should not depend on the bias voltage and test frequency. The larger capacitance values measured for one of the polarities and low test frequency resembles the well known CV characteristics of MOS structures \cite{MOSCap}. For MOS measurements, mobile surface charges determine the test results. They come from the leakage current, with the relevant polarity determined by the bias voltage. In particular, the capacitance measured in the inversion regime has a strong dependence on the test frequency. For the smaller frequency the larger capacitance is measured due to the surface charges motion. Our results may indicate the presence of the surface charges in the un-metallized areas on the top surface for the positive bias polarity.

A naive capacitance calculation can be done for the parallel plate model. An ideal capacitor, containing InP with a dielectric constant of 12.4 and 2 x 2 mm$^{2}$ conducting plates would have a capacitance of 1.25 pF. The closest measured analogue is the grounded GR high frequency CV measurement where the grounded GR should reduce edge effects. From the solid lines in the high-frequency CV measurements in Figure \ref{fig:cv_1mhz}, grounded GR high frequency capacitance is on average 1.8 pF. By contrast, floating GR high frequency capacitance is on average 2.0 pF. The discrepancy between measured capacitance and naive ideal capacitance can be attributed to the edge field distribution. The area of the backplane metallization is 25 $mm^2$, which is significantly larger than the area of the top pad of 4 $mm^2$. Therefore, a significant capacitance increase, compared to the equal-plate approximation, can be expected. The asymmetric architecture of devices, with a 25 mm$^2$ backplate and 4 mm$^2$ central pad, non-negligible edge effects are expected. 

The CV configuration should be made clear to properly contextualize grounded and floating GR. In all CV measurements, the AC wave from the LCR meter was only coupled to the central top pad and the backplane. In cases of grounded GR, the guard ring was connected to the constant ground in this configuration. This was equivalent to shielding the edge field at the periphery of the capacitor, resulting in a reduction of the measured capacitance.

\subsection{Electrical Tests Summary}
\label{electricalSummary}

Given the ohmic to non-ohmic behavior change at +350 V (Section \ref{subsectionIVmeas}), equivalently 10 kV/cm, may imply a transition point at this field strength, IV properties will be reported at this operating voltage. With only three high-current devices on record, high-current standard deviation (SD) and relative standard deviation (RSD, taken as SD divided by current magnitude) are not reported due to low statistics.

\begin{table}
\centering
\caption{Averaged leakage currents at +350 V and breakdown voltage. The results are shown in cases when only the central pad was grounded and when the guard ring was additionally grounded.}
\begin{tabular}{lrrrr}
\hline
Device type & \textbf{I$\mathrm{_{leak}}$(pad) [$\mu$A]} & \textbf{I$\mathrm{_{leak}}$(pad + GR) [$\mu$A]}  & \textbf{V$\mathrm{_{bd}}$(pad) [V]} \tabularnewline
\hline
Single-pad, typical       & 25.94 & 32.74 & 1010. \tabularnewline
Single-pad, high-current  & 29.12 & 37.38 & n.a. \tabularnewline
\hline
\end{tabular}
\label{IV_table}
\end{table}

\begin{table}
\centering
\caption{Measured leakage current standard deviations (SD) and relative standard deviations (RSD, taken as SD divided by leakage current) at +350 V.}
\begin{tabular}{lrr}
\hline
Device type & \textbf{I$\mathrm{_{RSD}}$(pad) [n.a.]} & \textbf{I$\mathrm{_{SD}}$(pad) [$\mu$A]} \tabularnewline
\hline
Single-pad, typical       &  0.092 & 2.40 \tabularnewline
\hline
\end{tabular}
\label{RSD_table}
\end{table}

\begin{table}
\centering
\caption{Measured capacitances at +20 V at 1 MHz.}
\begin{tabular}{lrr}
\hline
Device type & \textbf{C(floating GR) [pF]}& \textbf{C(grounded GR) [pF]} \tabularnewline
\hline
Single-pad, typical       &  2.09 & 1.76 \tabularnewline
Single-pad, high-current  & 2.08 & 1.83 \tabularnewline
\hline
\end{tabular}
\label{CV_table}
\end{table}

\section{Tests with micro-focused X-ray beams}
\label{sectionXrays}
The response of InP detectors to ionizing radiation was studied using micro-focused x-ray beams at the Canadian Light Source (CLS) and at the Diamond Light Source (DLS). These facilities operate monochromatic beams, which result in constant ioniazation amounts in a device under test (DUT) through the photoelectric effect. The electrons and holes from the DUT's active region were collected on its electrodes and read out as the current in the high-voltage power supply. 

Both facilities offer DUT spatial scanning in the beam, allowing the study of the active region's extent and uniformity for each sample \cite{Poley_2022}. Such tests have been done in the past for conventional silicon devices \cite{POLEY2020164509, SANTPUR2021164665}. For conventional silicon diodes, the studies showed a very high spatial uniformity.

\begin{figure}
    \centering
    \includegraphics[scale=0.3]{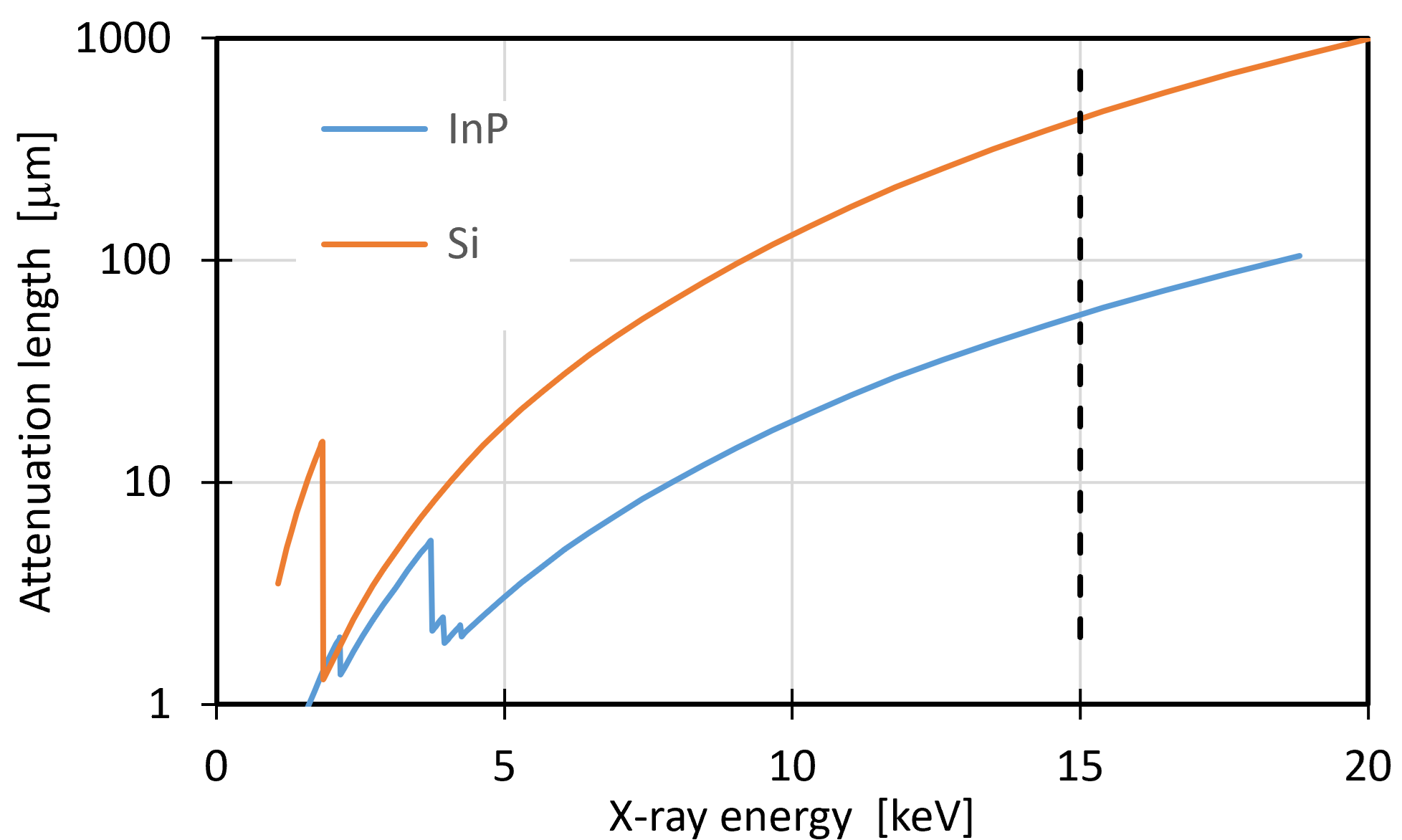}
    \caption{Absorption lengths in Si and InP detectors as a function of X-ray beam Energy. Beams at both DLS and CLS operated at 15 keV, denoted as a dashed vertical line.}
    \label{fig:Xray_absorption}
\end{figure}

Devices were translated horizontally and vertically across the beam to assess the active area of the sensor, with a steady interval of time between each movement step ($T_{step}$) (Table~\ref{scan_param_table}).  The active area is defined as the set of locations in the 2-dimensional area scan that exhibit a photocurrent response when stimulated by x-rays. This 2-dimensional projection of the 3-dimensional photoresponsive volume is dependent on the characteristic depth of InP:Fe bulk stimulated by x-rays, $d_{\mathrm{InP}}$, which is given by:
\begin{equation}
d_{\mathrm{InP}}=\frac{1}{\rho k_{eff}}
\end{equation}
where $\rho$ is the material density (4.81 g/cm for InP) and $k_{eff}$ is the mass attenuation coefficient.  For a composite material like InP, $k_{eff}$ is approximated by: 
\begin{equation}
k_{eff} \approx \sum_{\mathrm{elements}} \frac{\omega_Z}{\lambda_Z}
\end{equation}
where $\omega_Z$ is the percent weight of element $Z$ in the composite material, and $\lambda_Z$ is the mass attenuation length of element $Z$ for incident particles of a given energy~\cite{Workman:2022ynf}. In these studies, 15 keV X-rays were used (Figure \ref{fig:Xray_absorption}).
By treating the percent weight of iron in InP:Fe as negligible, the mass attenuation coefficient of InP:Fe ($k_{\mathrm{InP}}$) can be calculated from indium and phosphorous alone.  Under this assumption, $k_{\mathrm{InP}}=\num{37.6}~\mathrm{cm}^2$/g, and therefore $d_{\mathrm{InP}}=55.2~\mu$m. This translates to 90\% of X-ray photons being absorbed within the initial 127~$\mu$m. 

The experimental setup at DLS differs from that at CLS on two fronts: heat management and beam focus. Due to weak thermal coupling between the board's thermocouple and devices, measured device temperatures at CLS could not be reliably trusted. In order to have better temperature control, samples for DLS were mounted on a thinner board featuring a metal layer inset, which facilitated heat transfer. The working thermocouple used at DLS is known to have provided reliable device temperature information from other studies. Beam properties also differ between facilities, with the less focused beam at CLS chosen to provide a more stable beam with larger flux (Table~\ref{beam_table}).

The total current through the shared power supply, biasing both devices in parallel, was measured.
The current jumps measured in low voltage line scans and area scans outlined geometry with similar dimensions to known device features, mainly the central pad. These edges determined the mapping of physical features to fixed locations in all x-ray scans. One example is in Figure~\ref{fig:CLS_D2_Area}, where both the guard rings and the central pad are clearly visible. 
The device voltages used in tests were defined in quasi-logarithmic steps: 0.25, 1, 2, and 5 V at CLS and -20, -2, 5, 10, 15, 20, 30, 40, 50, 60, and 70 V at DLS. Two main types of scans are defined (line and area), differentiated only by the length of their step sizes in the $x$ and $y$ directions, listed in Table~\ref{scan_param_table}. Additional 400 V line scans, with different beam scan parameters, were taken at DLS to study persistent currents visible early on during data collection.

\begin{table}
\centering
\caption{Test Beam Scan Parameters}
\begin{tabular}{llrrrr}
\hline
Facility & Scan Type & $x_{step}$ [mm] & $y_{step}$ [mm] & $T_{step} [s]$ & Voltage Range [V]\tabularnewline
\hline
CLS & Line & 0.1 & 0.5 & 6 & 0.25 to 5\tabularnewline
CLS & Area & 0.075 & 0.075 & 6 & 2, 20\tabularnewline
DLS & 400V Line & 0.2 & n.a. & 7 & 400\tabularnewline
DLS & Line & 0.1 & 0.5 & 7 & -20 to 70\tabularnewline
DLS & Area  & 0.1 & 0.1 & 7 & 2, 20\tabularnewline
\hline
\end{tabular}
\label{scan_param_table}
\end{table}

\begin{table}
\centering
\caption{Beam Line Properties}
\begin{tabular}{lrrrrl}
\hline
Facility & Beam Energy & Beam Diameter & Operational & Beam Flux & Sample support\tabularnewline
&[keV] &[$\mu$m] &Temperature [C]& [ph/s]& \tabularnewline
\hline
CLS  & 15 & 40 & 2.5 & \SI{1.25e11}{} & Thick PCB\tabularnewline
DLS  & 15 & 2 & -15 & \SI{1.18e8}{}& Thin PCB\tabularnewline
&&&&&with enhanced\tabularnewline
&&&&&thermal control\tabularnewline
\hline
\end{tabular}
\label{beam_table}
\end{table}

\subsection{CLS Test Beam}
\label{subsectionCLS}

The board used in CLS contained two devices, labeled as Device 1 and Device 2 (Fig.~\ref{fig:CLS_board}). Voltage was applied to the backside conductors while both central pads were grounded. The guard rings around each central pad were left floating. While not realized at the time, Device 1 was a high-current device from the set described in Section \ref{subsectionIVuniform}.

\begin{figure}
    \centering
    \includegraphics[angle=270,width=\textwidth]{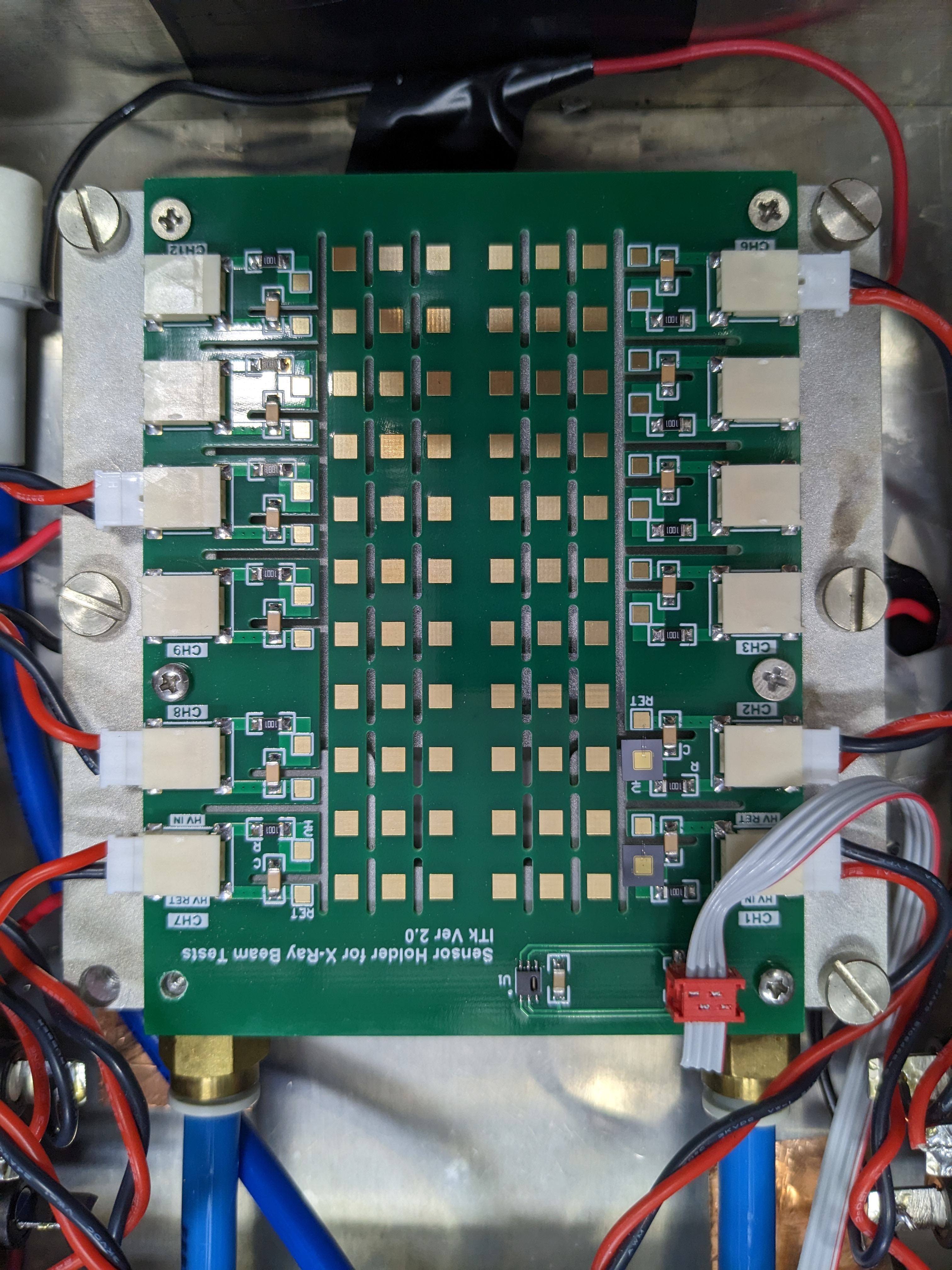}
    \caption{Photo of the board used and mounted on the CLS stage.}
    \label{fig:CLS_board_full}
\end{figure}

\begin{figure}
    \centering
    \includegraphics[width=\textwidth]{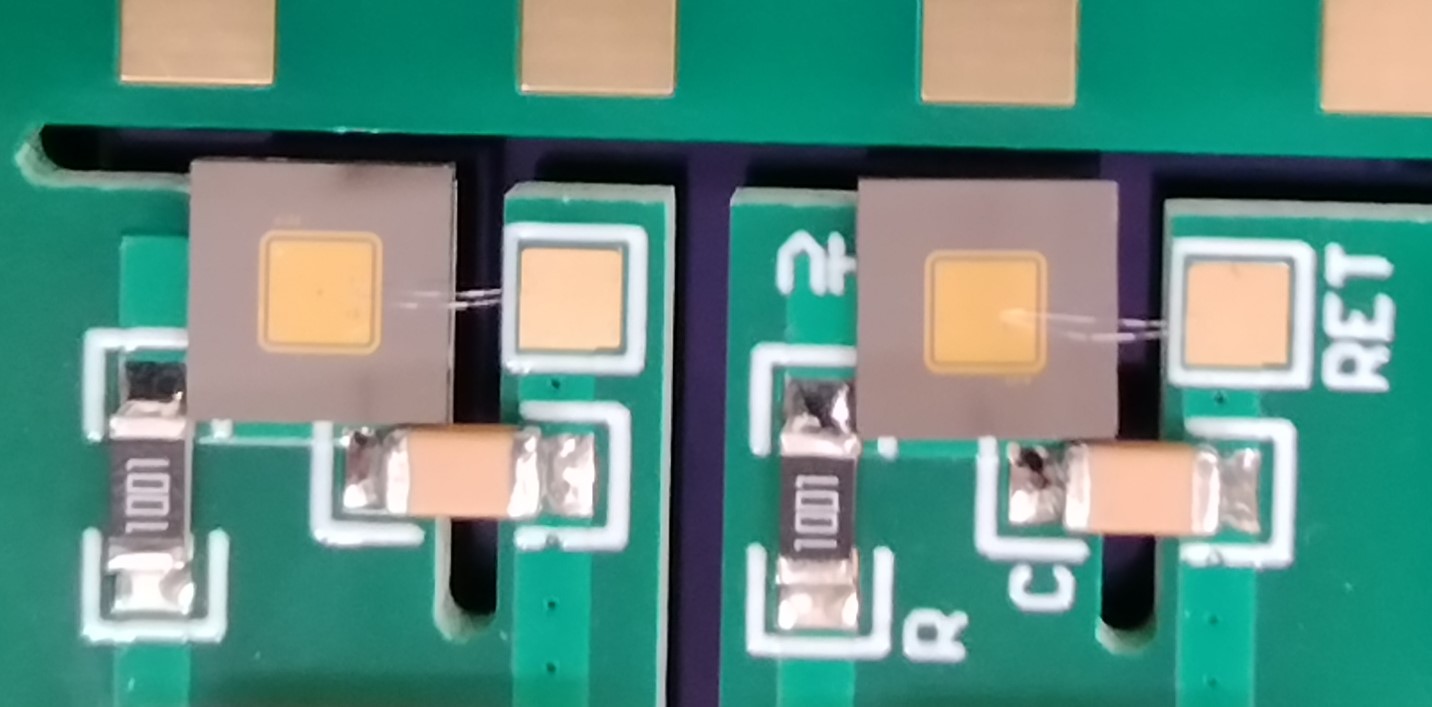}
    \caption{Photo of devices on CLS board, zoomed in from Figure \ref{fig:CLS_board_full}, oriented to match central pad area scan locations in CLS xy-coordinates.  Device 2 (A74$\_$H) is on the left, Device 1 (A73$\_$NH) is on the right.}
    \label{fig:CLS_board}
\end{figure}

\subsubsection{Device Map}
\label{subsubsectionCLSArea}

\begin{figure}

    \begin{subfigure}[b]{0.5\textwidth}
        \centering
        \includegraphics[width=1\textwidth]{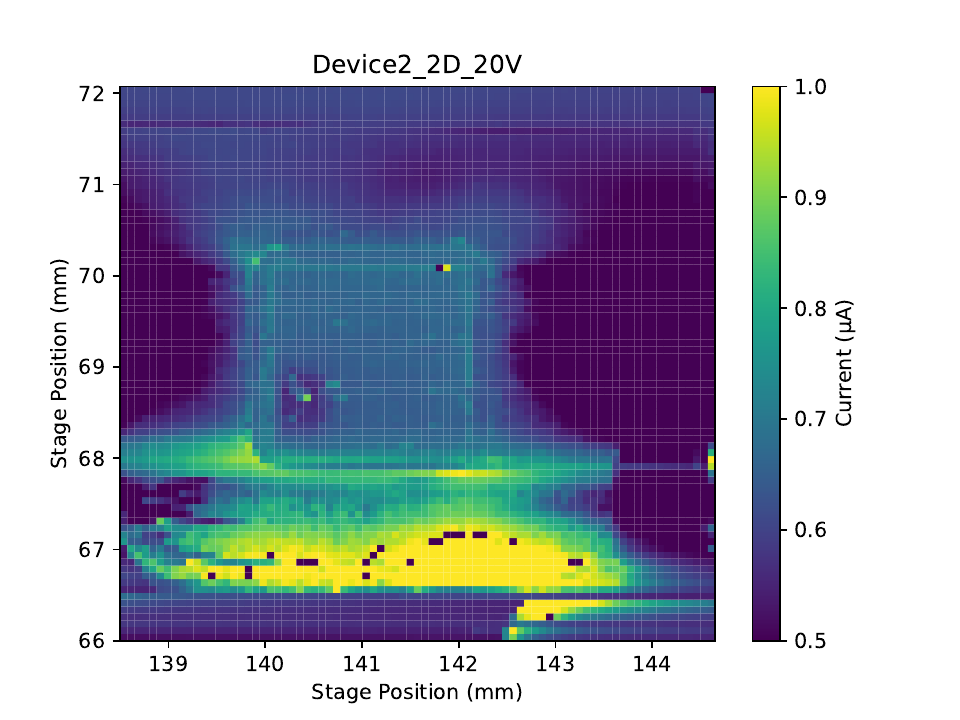}
        \caption{Device 2: 20 V}
        \label{fig:CLS_D2_Area}
    \end{subfigure}
    \begin{subfigure}[b]{0.5\textwidth}
        \centering
        \includegraphics[width=1\textwidth]{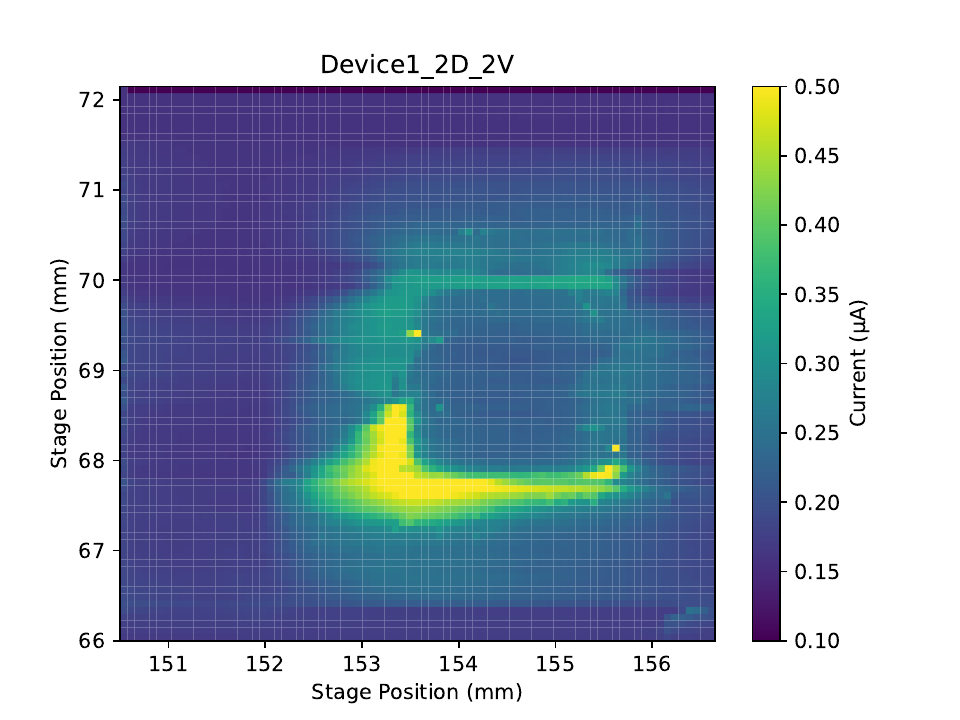}
        \caption{Device 1: 2 V}
        \label{fig:CLS_D1_Area}
    \end{subfigure}
    \caption{Area scan results from the CLS beam test.  The orientation of the devices in area scans matches the orientation of the CLS board shown in Fig.~\ref{fig:CLS_board}.  Device 2 (left) is scanned with a bias voltage of 2 V, while Device 1 (right) is scanned at 20V. Proper orientation was identified via x-ray scattering off of the CLS board capacitor, visible both in Figure \ref{fig:CLS_board} and in the bottom right hand corner of Figure (a). Dead pixels and flare-up are also visible in the bottom half of Device 2. Dead pixels arise from corrupted data entries during data taking.}
    \label{fig:CLS_Area}
\end{figure}

The only time-efficient way to precisely align the CLS precision stage coordinates of measured data to the positions of physical features was through an area scan of both devices. The alternative solution of taking a large area scan of all devices and beyond would take too much beam time for the same amount of information. Area scans detail the total current over a wide area of stage positions with high granularity in both x and y coordinates (Table \ref{scan_param_table}). The distance between devices in the x-axis and the capacitor visible in the bottom right of the Device 2 area scan were especially useful in determining the correlation between scan position and physical device position. 

Area scans were also used to reveal unexpected behavior that could affect the assessment of photocurrent uniformity. The region from $y = 66.5$ to $67.5$ mm in Device 2 in Figure \ref{fig:CLS_D2_Area} exhibits high currents. It corresponds to the device periphery and is uncorrelated to any other physical device features seen in Figure \ref{fig:CLS_board}. Since we are primarily interested in the active area around the central pad, line scans along the horizontal direction avoid this "flare-up" region by design, and uniformity assessment did not need to account for this unexpected behavior.

\subsubsection{Background Subtraction}
\label{subsubsectionCLSLine}

\begin{figure}
    \begin{subfigure}[b]{0.5\textwidth}
        \centering
        \includegraphics[width=\textwidth]{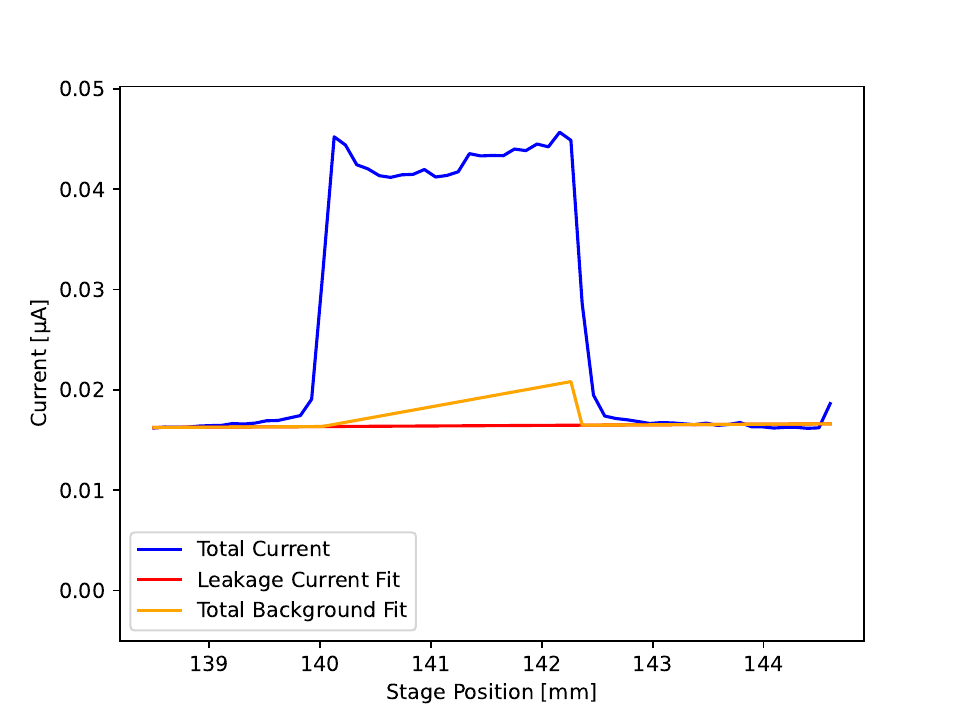}
        \caption{Defining Background Fit
        \\
        \\- Blue Line: The total raw current for constant y-position.
        \\- Red Line: A linear fit to non-photocurrent regions serves as the baseline leakage current.
        \\- Orange Line: A linear fit to the central pad region, accounting for photocurrent-dependent background.\\}
        \label{fig:CLS_Examp1}
    \end{subfigure}
    \hspace{1em}
    \begin{subfigure}[b]{0.5\textwidth}
        \centering
        \includegraphics[width=\textwidth]{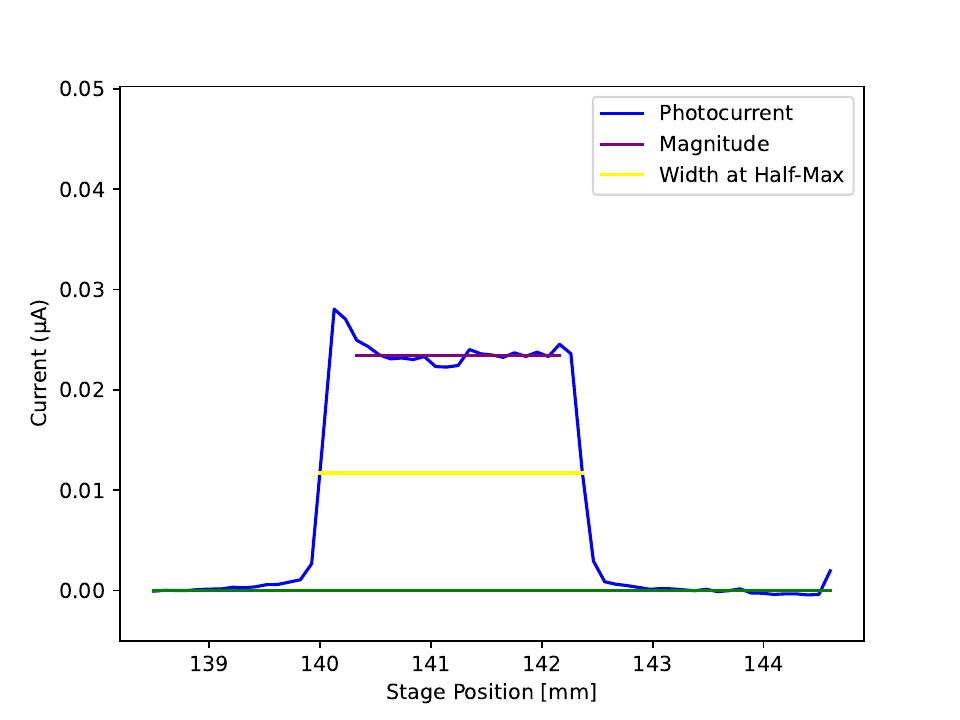}
        \caption{Subtracting Background Fit
        \\
        \\- Blue Line: Photocurrent was derived as the total current minus the total background.
        \\- Purple Line: The average photocurrent in the central pad region defines the photocurrent magnitude.
        \\- Yellow Line: Full Width at Half-Max was then defined as the width of the photocurrent region at half photocurrent magnitude.}
        \label{fig:CLS_Examp2}
    \end{subfigure}
    \caption{Line Scans of Device 2 at Line 0 with an applied voltage of 1 V, detailing the process of extracting photocurrent.}
\end{figure}

We model the measured total current as a combination of three different contributions: the photocurrent from the x-ray beam, an overall leakage current, and a photocurrent-induced background (Figure \ref{fig:CLS_Examp1}).

In order to measure the actual photocurrent, we first subtract the overall leakage current. The boundaries of regions absent of any photocurrent in all line scans were determined empirically. Finding these boundaries empirically was possible because of the high photocurrent to noise ratio visible in many line scans (Figure \ref{fig:CLS_Examp1}). Where there is no photocurrent, the overall leakage current was modelled by a linear fit, accounting for both a static leakage current and any time-dependent changes that would be difficult to accurately predict, but are to first-order linear as a function of time (i.e. leakage current temperature-dependence, power supply drift, etc). In regions with photocurrent, the overall leakage current is modelled as a linear extrapolation between the boundaries of non-photocurrent regions. Once the overall leakage current is defined throughout the line scan and subtracted out of the total current measured, the photocurrent-induced background should also be removed.

The x-ray beam and metallic surface of the central pad can result in non-photocurrent changes in the total current as a function of beam position (i.e. x-ray beam deflection off of the central pad surface and heating of the device from x-ray beam exposure in the active area). To account for this, after the overall background subtraction, a photocurrent-induced background is modelled as a linear fit confined to the central pad region. After both the photocurrent-induced background and overall leakage current are subtracted from the total current, the measured photocurrent remains. 

\begin{figure}
    \centering
    \includegraphics[width=\textwidth]{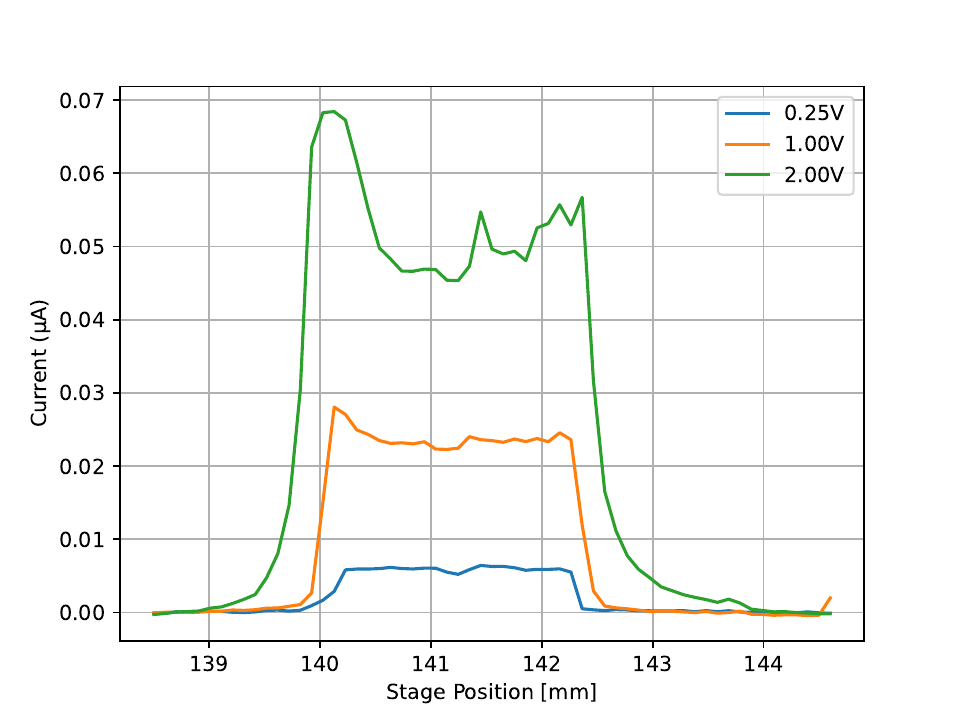}
    \caption{Photocurrent at Device 2, Line 0 from 0.25 to 2 V.}
    \label{fig:CLS_I_Stack}
\end{figure}

The beam intensity fluctuates over time, which introduced uncertainty in determining uniformity purely from physical features of the device. This uncertainty was accounted for by dividing the photocurrent by beam intensity, normalized by the beam intensity order of magnitude, leaving a scaled current whose active area can be assessed (Figures \ref{fig:CLS_Examp2}, \ref{fig:CLS_I_Stack}). This beam intensity derived scaling factor is numerically close to one, visible in the photocurrent maintaining the same order of magnitude from Figure \ref{fig:CLS_Examp1} to \ref{fig:CLS_Examp2}.

\begin{figure}[!tbp]
  \centering
  \begin{minipage}[b]{0.45\textwidth}
    \includegraphics[width=\textwidth]{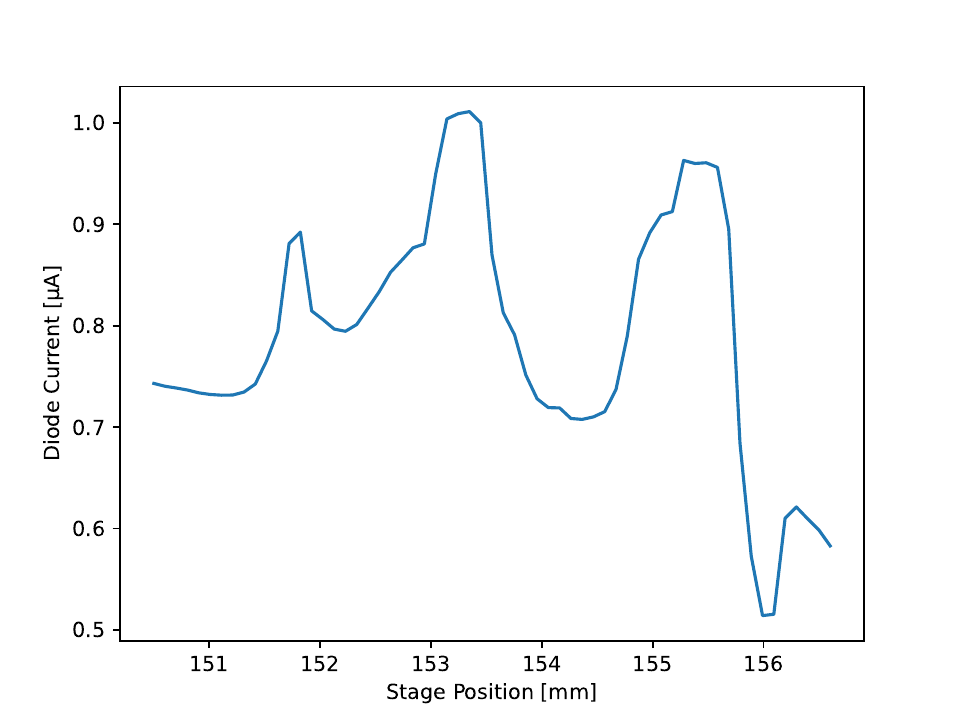}
    \caption{Total current measured in Device 1, Line 1 at 5 V. Unexplained spikes in current are visible near 152 and 156 mm, making background fits difficult to calculate.}
    \label{fig:CLS_5V}
  \end{minipage}
  \hfill
  \begin{minipage}[b]{0.45\textwidth}
    \includegraphics[width=\textwidth]{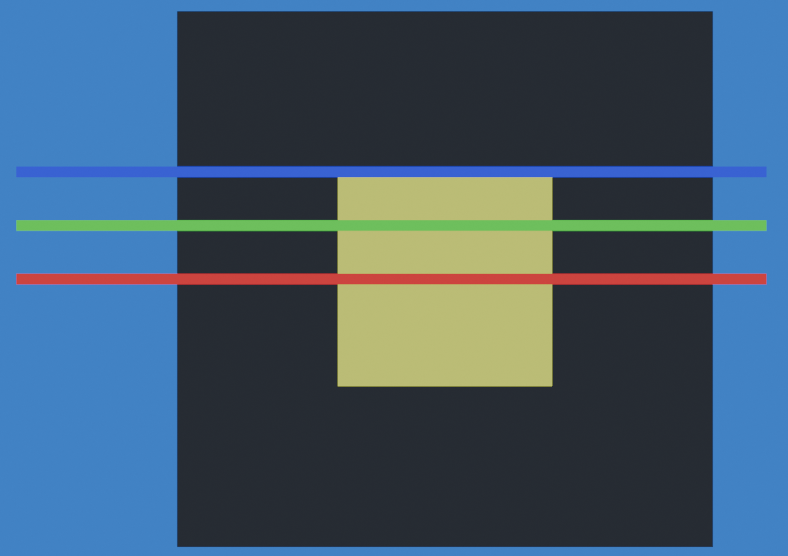}
    \caption{Lines 0, 1, and 2 (red, green, and blue lines) were positioned to cross through the center, off-center, and edges of both devices based on area scan information. The device is represented as a black square with a yellow central pad, with the guard ring omitted for visual clarity.}
    \label{fig:CLS_Line_Rep}
  \end{minipage}
\end{figure}

\subsubsection{Results}
\label{subsubsectionCLSLineResults}

\begin{figure}
    \begin{subfigure}[b]{0.5\textwidth}
        \centering
        \includegraphics[width=\textwidth]{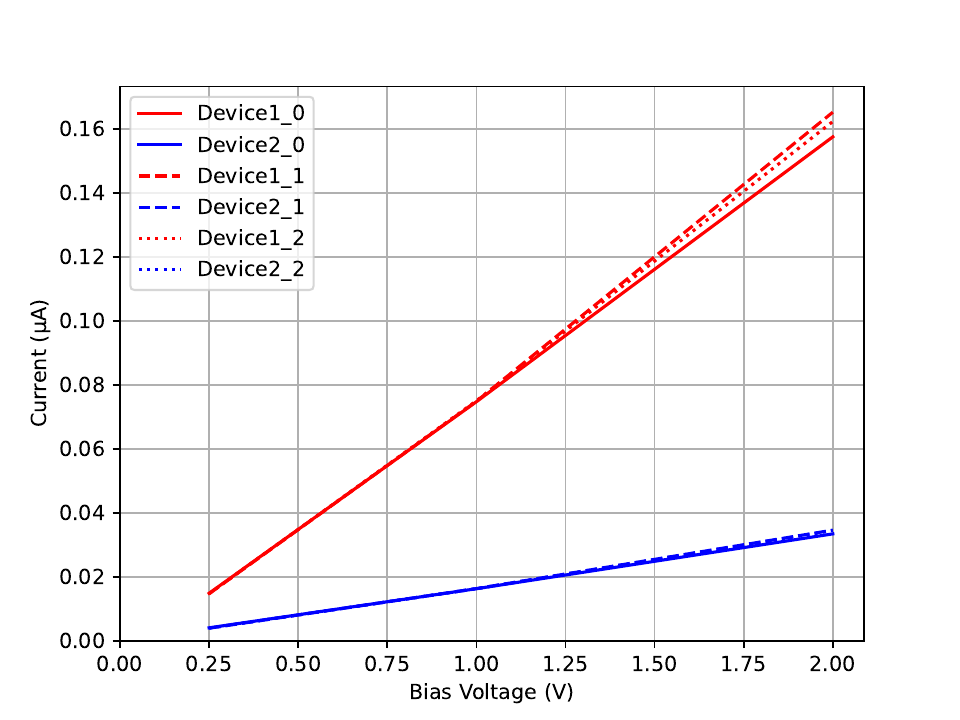}
        \caption{Leakage Current vs Voltage}
        \label{fig:CLS_leakage}
    \end{subfigure}
    \begin{subfigure}[b]{0.5\textwidth}
        \centering
        \includegraphics[width=\textwidth]{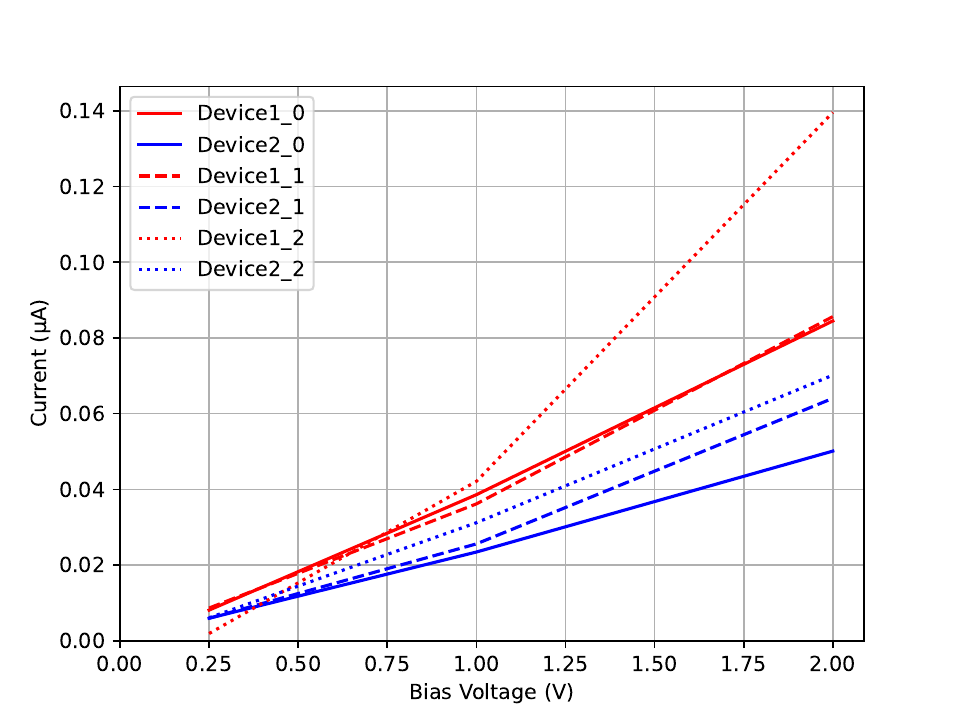}
        \caption{Magnitude vs Voltage}
        \label{fig:CLS_photocurrent}
    \end{subfigure}
    \begin{subfigure}[b]{0.5\textwidth}
        \centering
        \includegraphics[width=\textwidth]{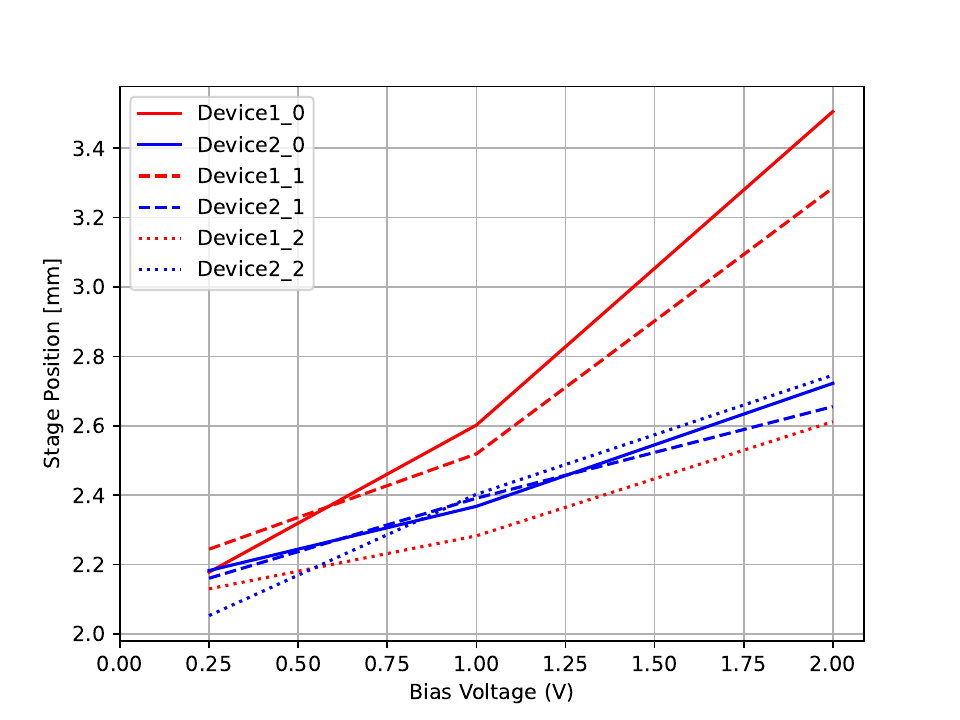}
        \caption{Active Area Width vs Voltage}
        \label{fig:CLS_areawidth}
    \end{subfigure}
    \begin{subfigure}[b]{0.5\textwidth}
        \centering
        \includegraphics[width=\textwidth]{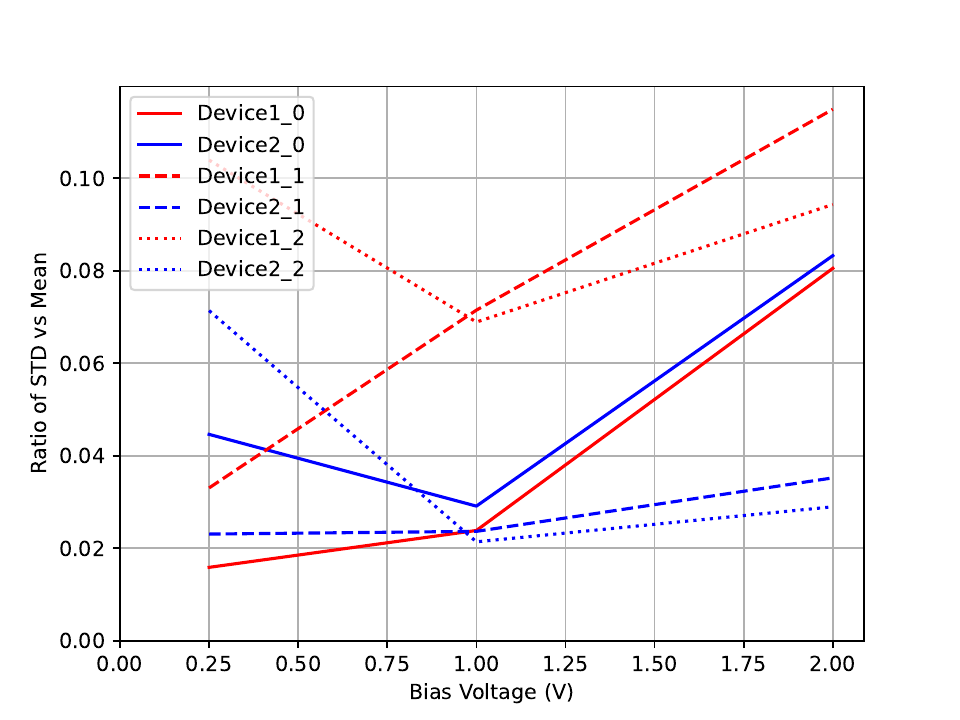}
        \caption{Relative Standard Deviation vs Voltage}
        \label{fig:CLS_rsd}
    \end{subfigure}
    \caption{Figure \ref{fig:CLS_leakage} shows the leakage current in photocurrent-absent regions at each voltage. Figures \ref{fig:CLS_photocurrent}-\ref{fig:CLS_rsd} detail photocurrent properties as a function of voltage. Data from devices 1 and 2 are shown as red and blue lines in these plots. Line positions 0, 1, and 2 correspond to solid, dashed, and dotted lines respectively.}
    \label{fig:CLS_line_results}
\end{figure}

A subset of CLS data, in ranges of position and voltage, will be assessed. Since calculating active area uniformity only requires photocurrent information under the central pad area, only horizontal scans which sweep over the central pad area were used: Lines 0, 1, and 2 (Figure \ref{fig:CLS_Line_Rep}). Summary plots omit voltages higher than 5 V due to an issue with background subtraction at higher voltages, stemming from low statistics in non-photocurrent regions visible in total current plots (Figure \ref{fig:CLS_5V}). The presence of unexplained current spikes in normally non-photocurrent regions muddles the reliability of background subtraction at higher voltages. Since temperature could not be accurately measured due to the board's unreliable thermocouple, temperature fluctuations could not be accounted for or ruled out as an explanation for these current spikes.

Before photocurrent magnitude, width, and relative standard deviation (RSD) could be assessed, they needed to be algorithmically defined. Photocurrent RSD was calculated within 1 mm of the central pad center, corresponding to the central pad region. Magnitude was defined as the average photocurrent in this same range. Width could then be evaluated as the distance across the photocurrent plateau at half of the photocurrent magnitude (Figure \ref{fig:CLS_Examp2}).  Each of these features was calculated for each Line in each line scan (Figure \ref{fig:CLS_line_results}).

Despite the large fluctuations observed at high voltages, key conclusions derived were useful for both improvements at the subsequent test beam run at DLS and for assessing InP response properties. 

Photocurrent RSD was taken as as a measure of intra-device uniformity. RSD values measured at CLS are lower than, though roughly on the same order as, those attributed to room temperature noise measured in \ref{subsectionIVuniform} (Figure \ref{fig:CLS_rsd}, \ref{fig:iv_rsd}). This implies signal amplitude uncertainties both from an unregulated-temperature environment and from bulk non-uniformity contribute at roughly the same order with low voltages: a 2-10 percent photocurrent variation within one standard deviation. Compared to other photocurrent attributes, RSD dependence on voltage or line position is less clear, given the narrow voltage range used. 

\begin{figure}
    \begin{subfigure}{\textwidth}
        \centering
        \includegraphics[width=0.6\textwidth]{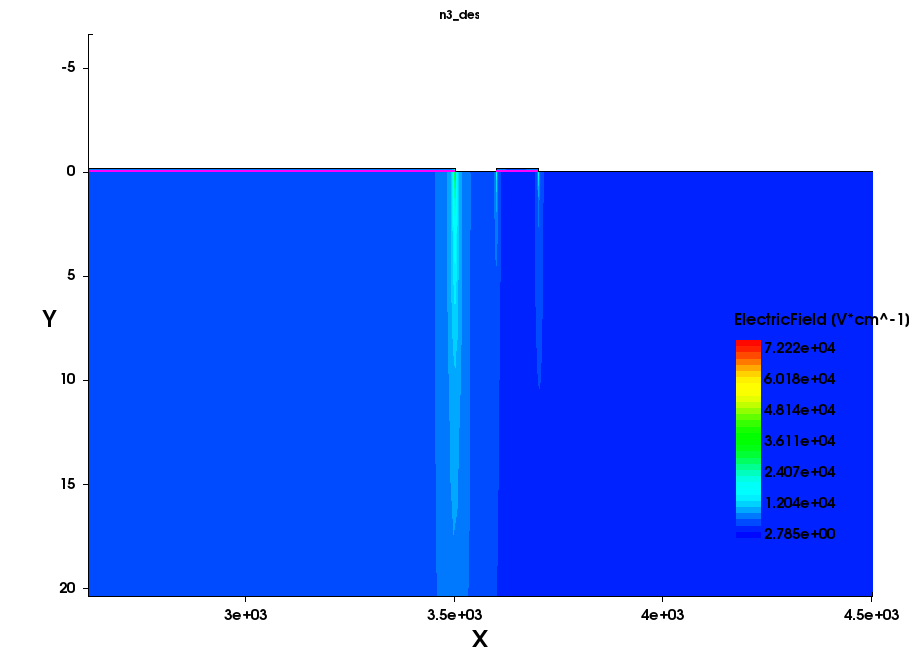}
        \caption{Electric Field Magnitude Edge Profile}
        \label{fig:EM_sim_sideview}
    \end{subfigure}
    \begin{subfigure}{\textwidth}
        \centering
        \includegraphics[width=0.6\textwidth]{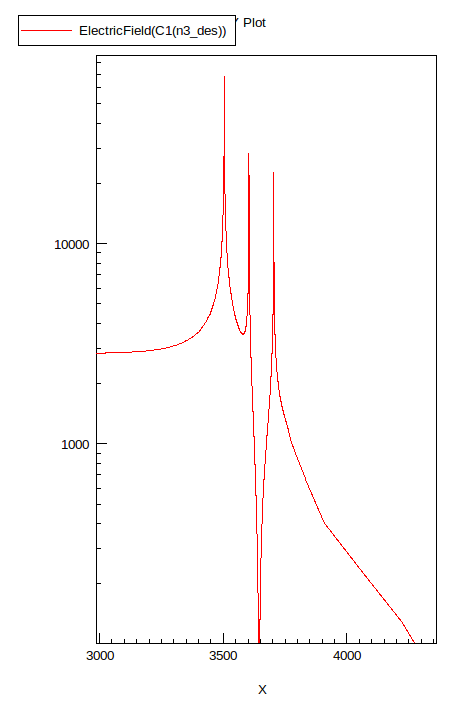}
        \caption{Device Surface Electric Field Magnitude}
        \label{fig:EM_sim_top_cut}
    \end{subfigure}
    \caption{Sentaurus 2D electric field simulation of a generic InP:Fe device, with the backplane set at 100 V, viewed as an edge profile (\ref{fig:EM_sim_sideview}) and surface field magnitude (\ref{fig:EM_sim_top_cut}) \cite{Sentaurus}. Color denotes electric field strength. A high electric field strength region was identified at the central pad edge (3500 microns) and guard ring edges (3600-3700 microns) both at the device bulk and surface.}
    \label{fig:EM_sim}
\end{figure}

Magnitude and width were measured to determine how the photoresponse strength and active area size change as a function of voltage and position (Figure \ref{fig:CLS_photocurrent}, \ref{fig:CLS_areawidth}). The width at the smallest voltage value corresponds to the distance between the outer edges of the guard ring. Both the magnitude and width increase as a function of bias voltage. We attribute this observation to faster charge carrier drift at higher fields, resulting in a smaller effective trapping. While the active area width does not appear to heavily depend on scan position, Lines further from the central pad center tended to have a higher magnitude, implying that photoresponse increases closer to the central pad border where there is a peak of the electric field due to the very thin sharp edge of the top metal (Figure \ref{fig:EM_sim}). For silicon devices, this issue is attenuated by the natural implant depth of about 1-2 mm. For InP devices, there was no special implantation, therefore the thin metal is likely responsible for the edge effects observed. At DLS, we circumvented the impact of edge effects on photocurent measurements by assessing magnitude and RSD in a narrower physical range in the active area. 

The inclusion of high-current device A73\_NH/Device 1 at CLS provided an opportunity to assess the behavior of high-current devices under an x-ray test beam. High leakage current at low voltages expected from high-current devices remains consistent at CLS (Figure \ref{fig:CLS_leakage}). Device 1 showed a higher photocurrent magnitude and width in comparison to Device 2, though the behavior of RSD in high-current devices compared to nominal devices could not be assessed from Figure \ref{fig:CLS_rsd}. This photocurrent behavior may hint that the differences found in high-current properties may relate to bulk properties.

\subsection{DLS Test Beam}
\label{subsectionDLS}

\begin{figure}
    \centering
    \includegraphics[width=\textwidth]{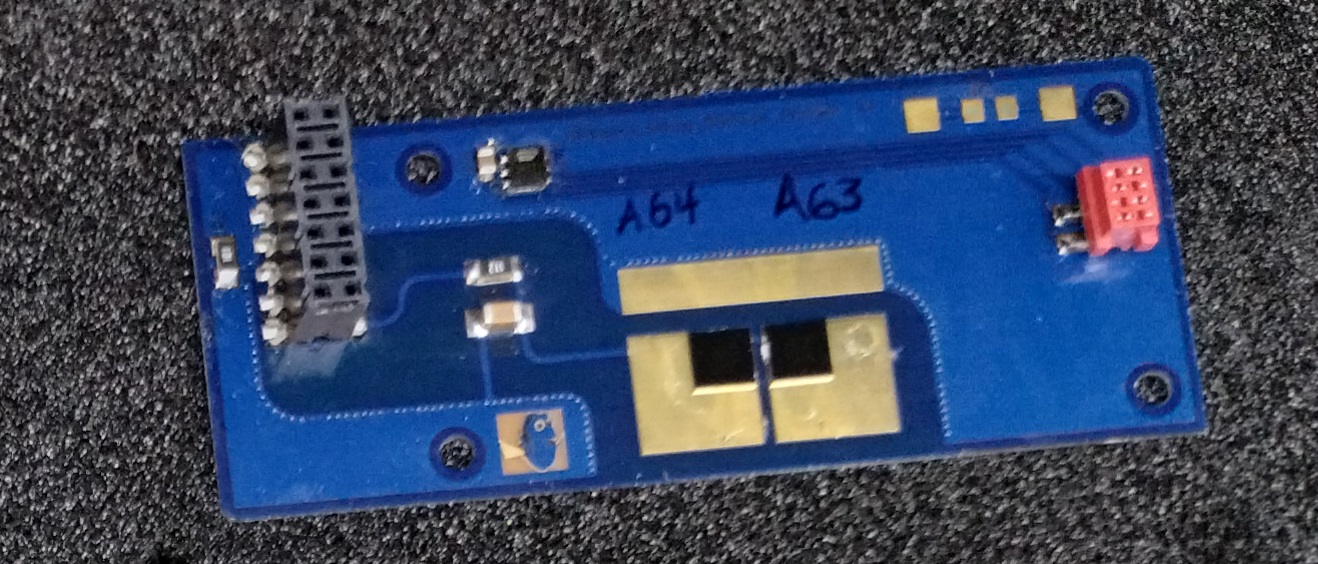}
    \caption{Visual of the board used at DLS with A64\_H (Device 2) and A63\_H (Device 1) mounted.}
    \label{fig:DLS_board}
\end{figure}

The DLS beam test offered an opportunity to iterate on the experimental setup from CLS. A different pair of devices from those at CLS were chosen for DLS: A64\_H and A63\_H (Figure \ref{fig:DLS_board}). Both devices are mounted on a board with better thermal readout and dissipation. Only sub-celsius changes in temperature were recorded across line scans.

The background subtraction technique detailed in \ref{subsubsectionCLSLine} is reused to extract the photocurrent from DLS data. Information on the parameters of the experimental setup can be found in Tables \ref{beam_table} and \ref{scan_param_table}.

To help visualize the y-positions of lines relative to device features, line y-positions 10.15, 10.65, and 11.15, and 11.65 mm will be referenced as below-center, center, above-center, and edge (Figure \ref{fig:DLS_colorcode}). Images will still retain the number-based marker for accurate referencing to Figure \ref{fig:DLS_area}.

\subsubsection{Device Map}
\label{subsubsectionDLSArea}

\begin{figure}
    \begin{subfigure}[b]{0.5\textwidth}
        \centering
        \includegraphics[width=\textwidth]{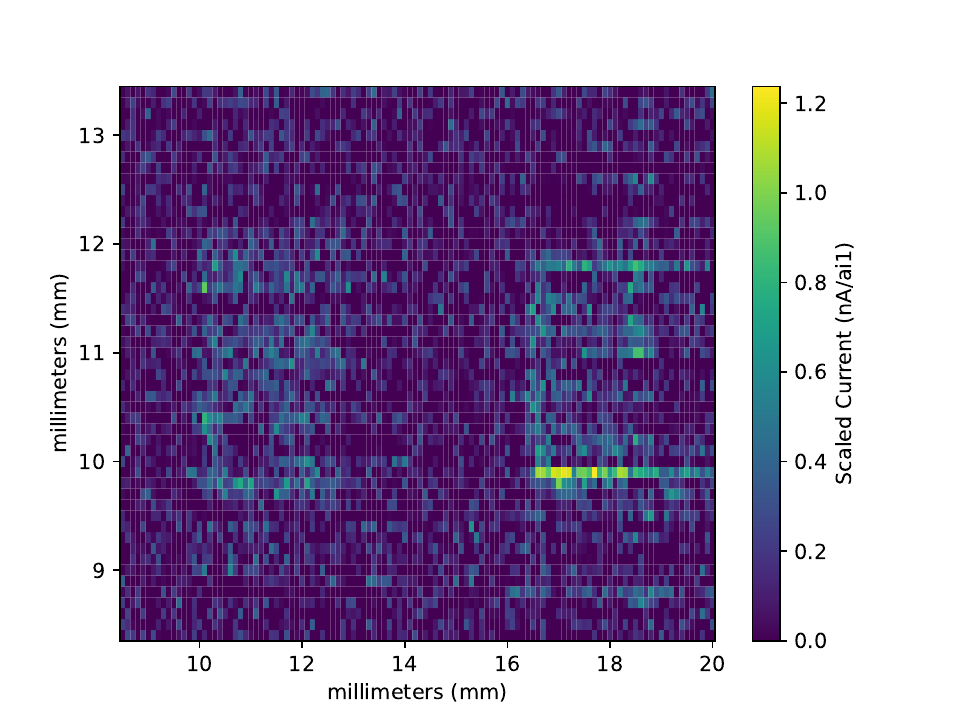}
        \caption{2 V}
        \label{fig:DLS_area2}
    \end{subfigure}
    \begin{subfigure}[b]{0.5\textwidth}
        \centering
        \includegraphics[width=\textwidth]{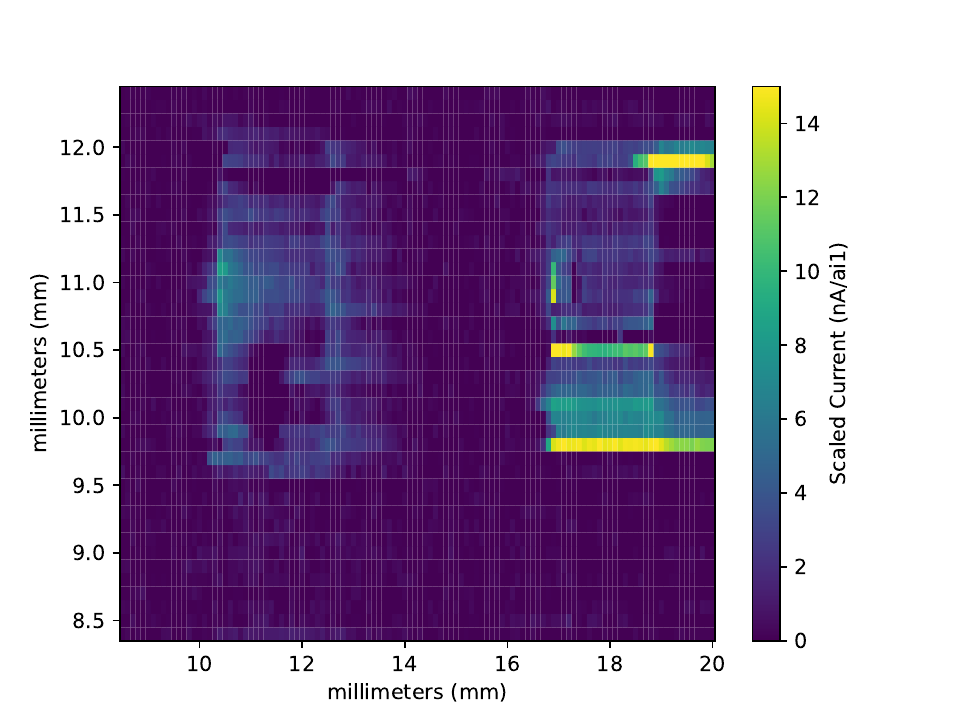}
        \caption{20 V}
        \label{fig:DLS_area20}
    \end{subfigure}
    \begin{subfigure}[b]{0.5\textwidth}
        \centering
        \includegraphics[angle=-90, width=\textwidth]{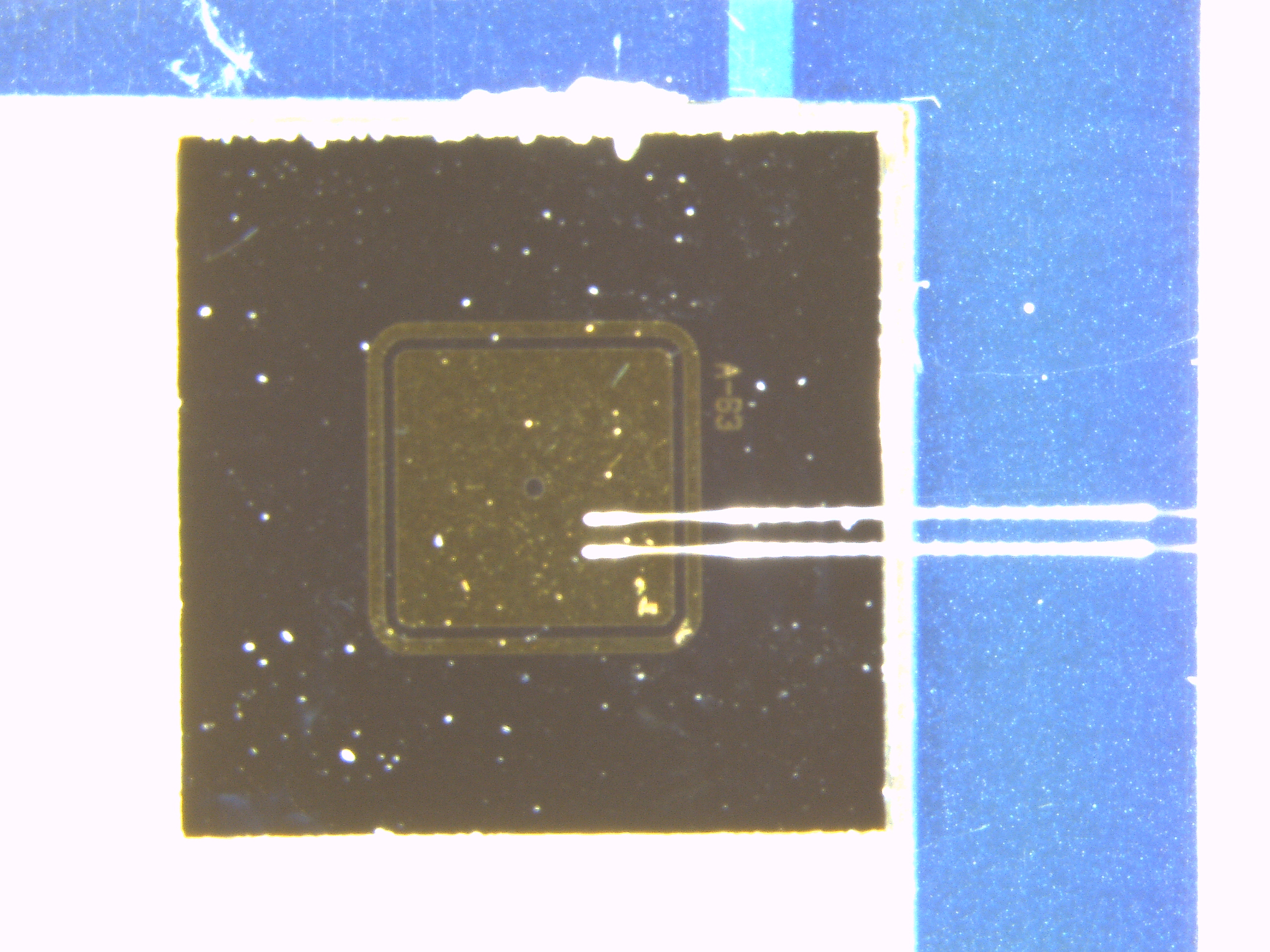}
        \caption{A63\_H -- Zoomed In}
        \label{fig:DLS_A63}
    \end{subfigure}
    \begin{subfigure}[b]{0.5\textwidth}
        \centering
        \includegraphics[angle=-90, width=\textwidth]{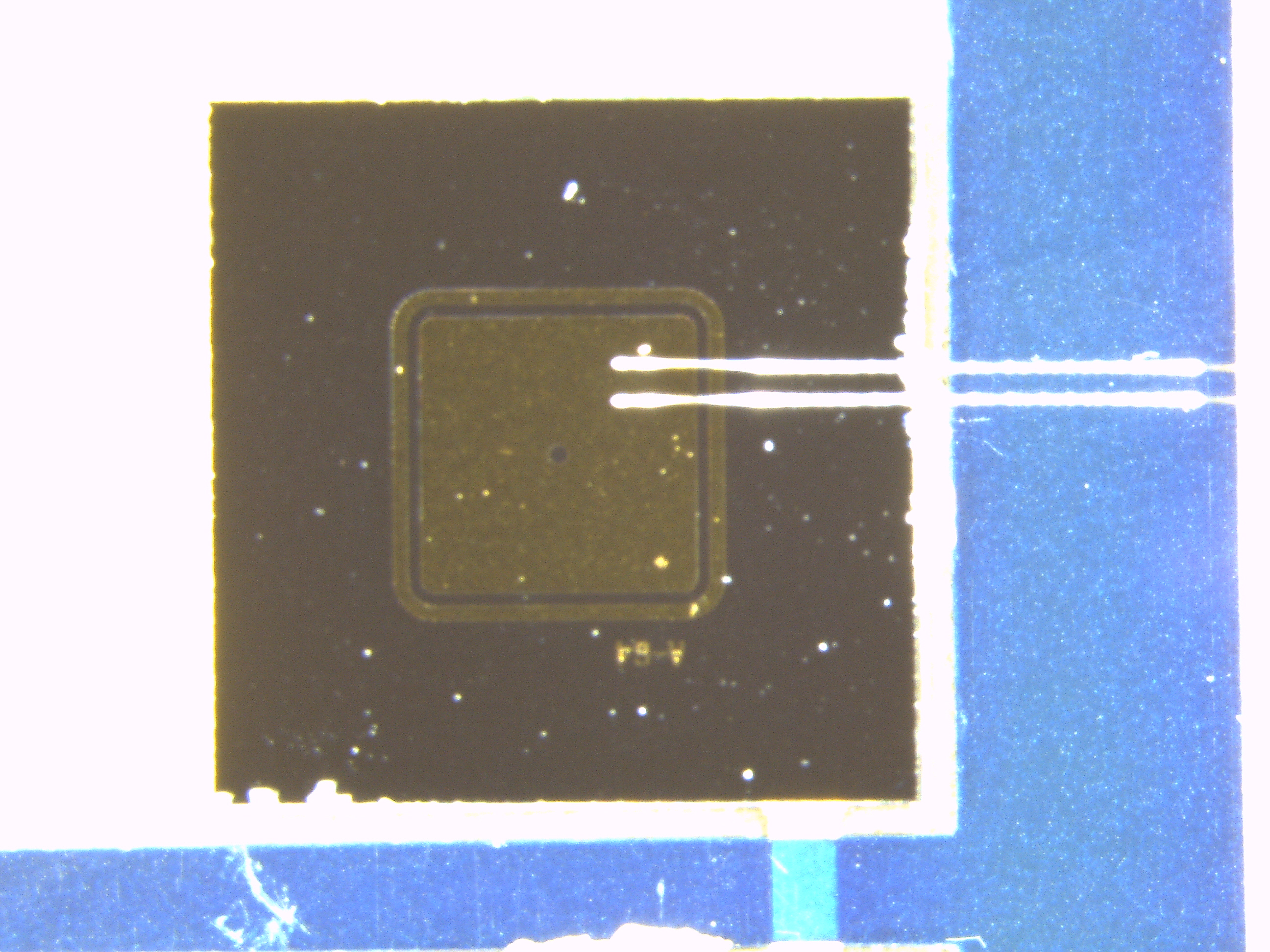}
        \caption{A64\_H -- Zoomed In}
        \label{fig:DLS_A64}
    \end{subfigure}
    
    \caption{Area scans at 2 V (a), 20 V (b), and visuals for device 1 (c) and device 2 (d) used at DLS in matched orientation.}
    \label{fig:DLS_area}
\end{figure}

\begin{figure}
    \centering
    \includegraphics[width=\textwidth]{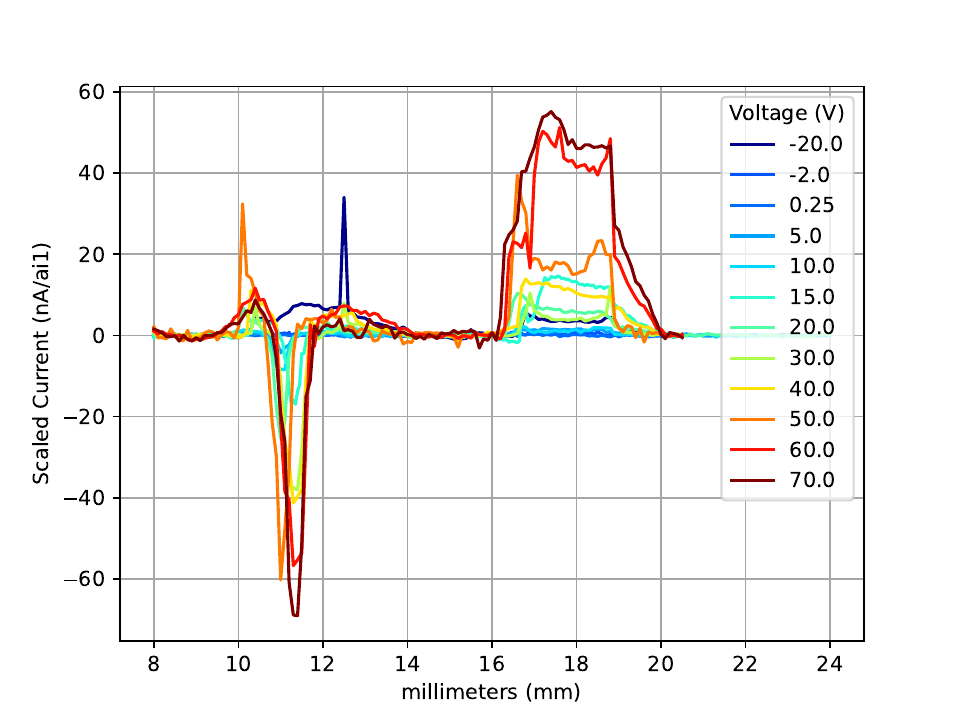}
    \caption{Scaled photocurrents (current divided by normalized beam flux) were calculated at the below-center line, with current polarity adjusted to show photocurrent magnitude. Device 1 is centered along the x-axis at x = 11.5 mm and Device 2 is centered at x = 17.8 mm. Dip current is visible in Device 1.}
    \label{fig:DLS_iy_10-15}
\end{figure}

\begin{figure}
    \begin{subfigure}[b]{0.5\textwidth}
        \centering
        \includegraphics[width=\textwidth]{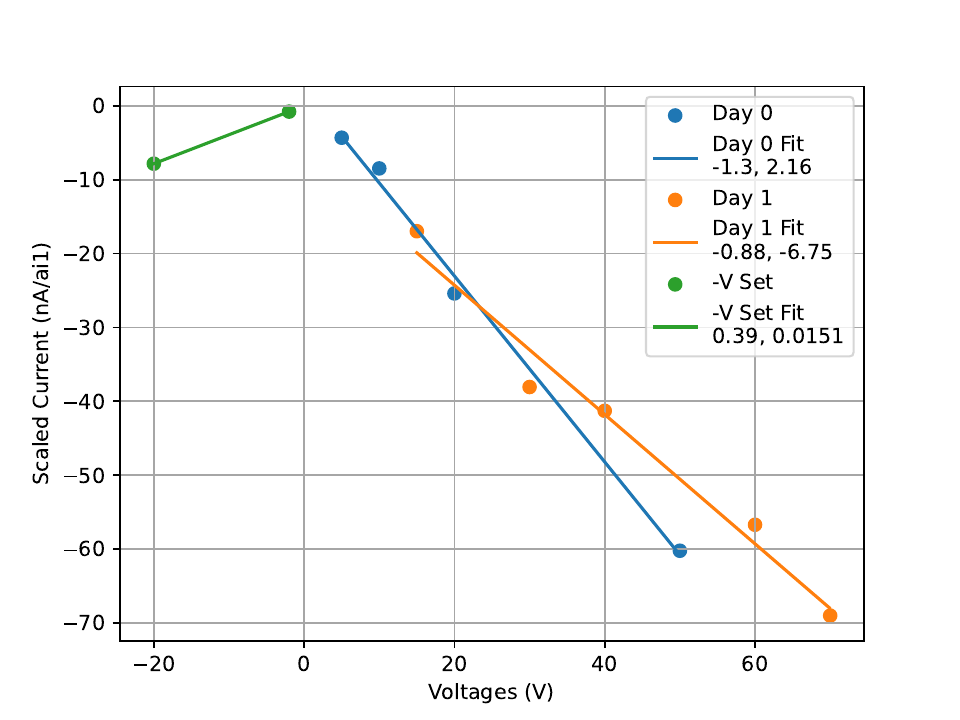}
        \caption{Dip Magnitude vs Voltage}
        \label{fig:DLS_dip_mag}
    \end{subfigure}
    \begin{subfigure}[b]{0.5\textwidth}
        \centering
        \includegraphics[width=\textwidth]{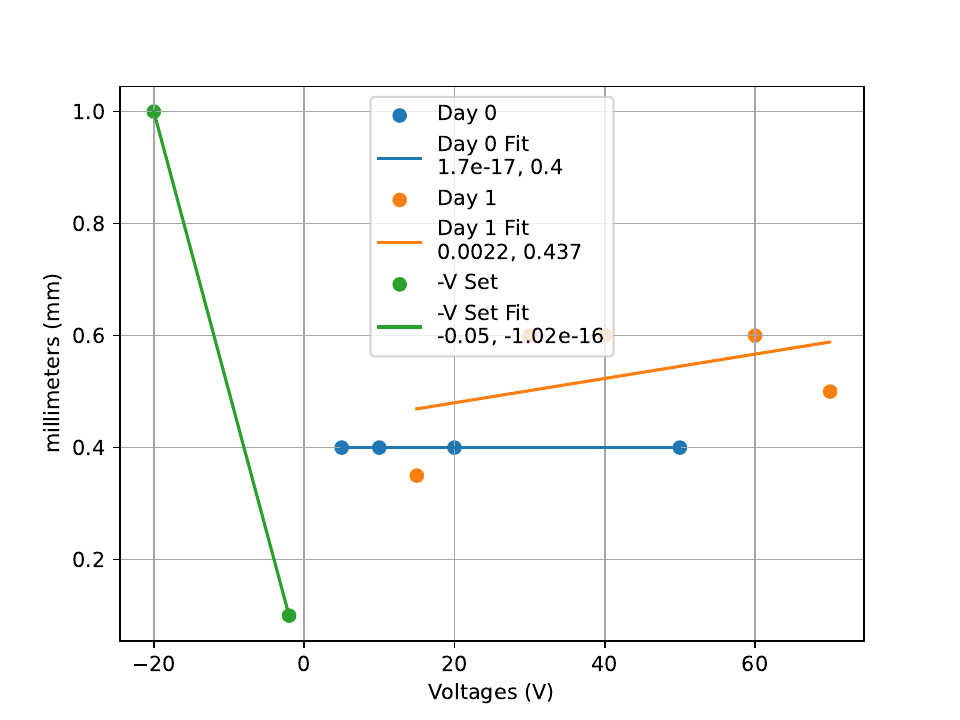}
        \caption{Dip Width vs Voltage}
        \label{fig:DLS_dip_width}
    \end{subfigure}
    
    \caption{Dip properties derived after background subtraction. The photocurrent maximum and FWHM (full-width half-max) of the dip current are represented as dip magnitude and width. Data is grouped and colored by measurement day, except for a negative voltage group defined separately to help with fitting. Linear fits to each group are also shown. Legend entries for linear fits detail the slope and y-axis offset for each fit (m, b). 0.25 V data was excluded due to high noise.}
    \label{fig:DLS_dip_stats}
\end{figure}

Area scans held at 2 and 20 V were performed at DLS to identify the position and orientation of devices with respect to total current features (Figure \ref{fig:DLS_area}). A63\_H and A64\_H will be referred to as Device 1 and Device 2 to help communicate the association between physical devices and photocurrent regions. Device 1 and Device 2 central pads correspond to the left and right photocurrents visible in the 2 V area scan (Figure \ref{fig:DLS_area2}). 

Beyond acting as a position reference, area scans were also used to identify abnormal features with high granularity. Higher beam currents obtained from the narrower beam diameter at DLS also served to increase the magnitude of observed photocurrent and any x-ray dependent abnormal phenomena. In comparison to CLS tests, flare-up was not observed in DLS area scans. Whether flare-up observed at CLS was due to unique non-uniformity at that specific device's periphery or not therefore remains unresolved. 

Centered at (11.3, 10.25) mm, a spatially-localized spike in photocurrent was identified on Device 1's central pad (Figures \ref{fig:DLS_area20}, \ref{fig:DLS_iy_10-15}). While the magnitude of this "dip" current, after background subtraction, increases linearly as a function of bias voltage, its width remains steady on each measurement day (Figure \ref{fig:DLS_dip_mag}, \ref{fig:DLS_dip_width}). Differences in dip current properties between Day 0 and Day 1 likely stem from a sub-mm offset in stage positions between days. With an approximate width of 0.5 mm, the area covered by the dip current is 0.2 mm$^2$, making up about 5\% of the Device 1 central pad area. 

The leading hypothesis for the cause of dip current is a crack in the bulk of Device 1, caused by a wirebonding process. The dip location matches the bondfeet placement, and wirebonding with InP devices proved difficult due to the thin central pad metallization, requiring a larger than usual force to make welds. Additionally, a bonding process applied to unpatterned InP surface indicated a possibility of the surface damage. Possible solutions to address this issue in future devices, such as thicker metallization, are under consideration. Lines containing the dip current (i.e. below-center and center lines on Device 1) will not be used in evaluating InP uniformity.

\subsubsection{Persistent Current}
\label{subsubsectionDLSpersistent}

\begin{figure}
    \begin{subfigure}[b]{0.5\textwidth}
        \centering
        \includegraphics[width=\textwidth]{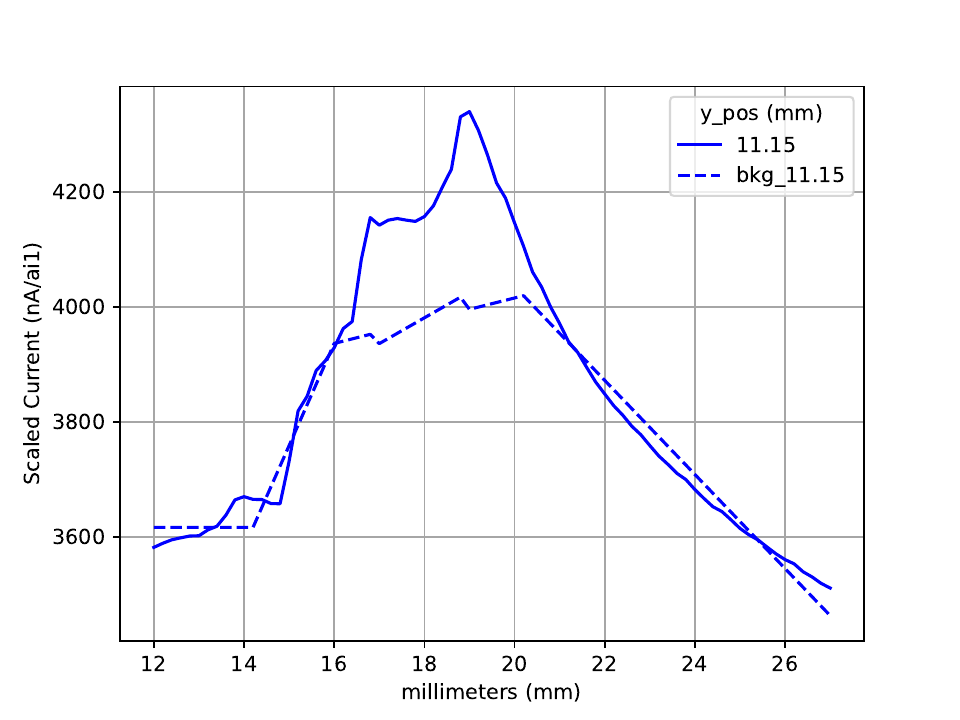}
        \caption{Full Beam Data over Stage Position}
        \label{fig:DLS_persistent_pos}
    \end{subfigure}
    \begin{subfigure}[b]{0.5\textwidth}
        \centering
        \includegraphics[width=\textwidth]{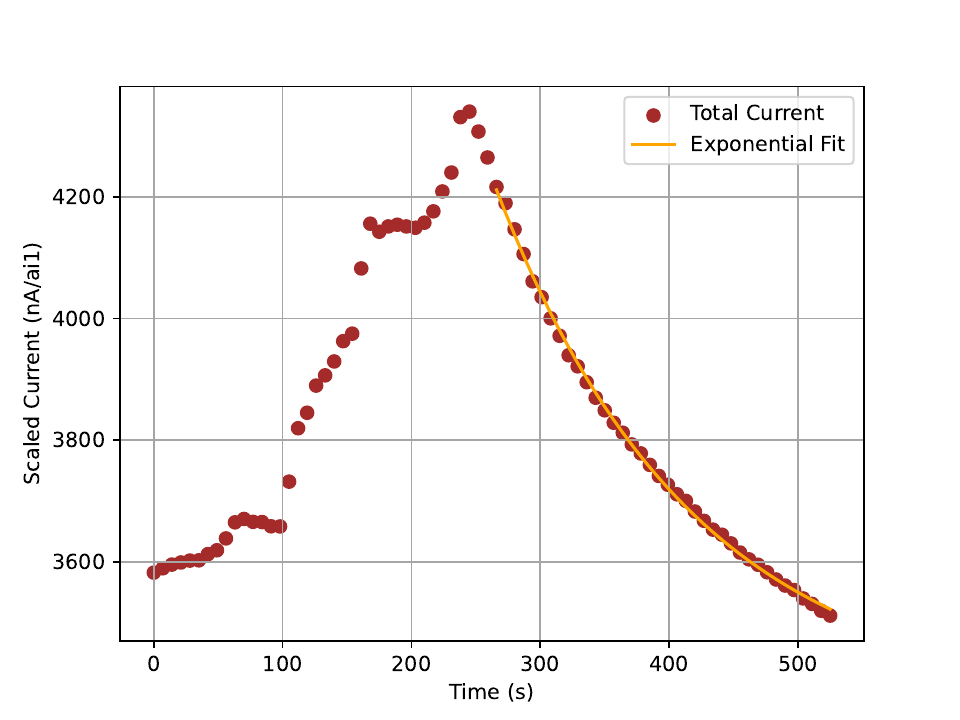}
        \caption{Full Beam with Data and Fit}
        \label{fig:DLS_persistent_full}
    \end{subfigure}
    \begin{subfigure}[b]{0.5\textwidth}
        \centering
        \includegraphics[width=\textwidth]{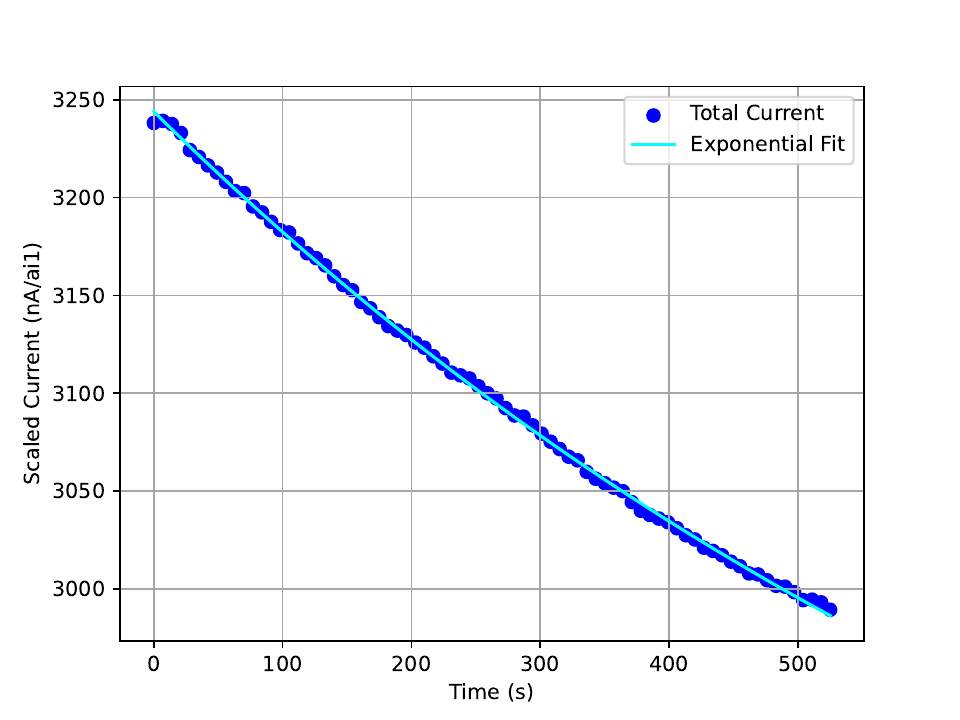}
        \caption{No Beam with Data and Fit}
        \label{fig:DLS_persistent_att}
    \end{subfigure}
    \begin{subfigure}[b]{0.5\textwidth}
        \centering
        \includegraphics[width=\textwidth]{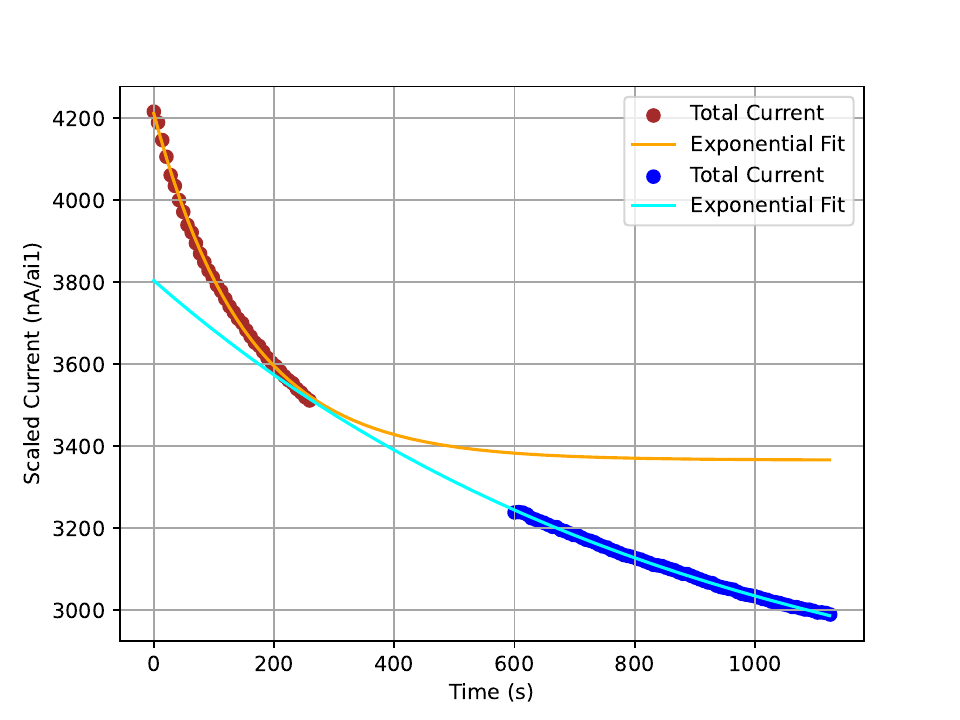}
        \caption{Sequential Full and No Beam with Data and Fits}
        \label{fig:DLS_persistent_all}
    \end{subfigure}
    
    \caption{Persistent current investigation using data taken at 400 V. A plot of the total current as a function of position is provided to show that 0.2 mm movement in x-position is mapped to 7 seconds of held beam position (Table \ref{scan_param_table}, Figures \ref{fig:DLS_persistent_pos}, \ref{fig:DLS_persistent_full}). Figures \ref{fig:DLS_persistent_full} and \ref{fig:DLS_persistent_att} show the full and no beam total current over time, alongside exponentially falling fits. Figure \ref{fig:DLS_persistent_all} plots data from both scans over time, with a gap of 341 seconds occurring between the end of the full beam scan and the start of the no beam scan.}
    \label{fig:DLS_persistent}
\end{figure}

An effort was made to identify high-voltage phenomena early by performing line scans at 400 V, operating well below the breakdown voltage observed in Section \ref{subsectionBreakdown}. From 400 V line scans, a measured excess current was observed to persist for several minutes after x-ray stimulation was halted. Larger step sizes were used to quickly assess this persistent current behavior, as its discovery occurred in the middle of allocated test beam time (Table \ref{scan_param_table}). Two line scans were performed at 400 V in sequence: the first with full beam flux through InP devices and the second without any beam flux (Figure \ref{fig:DLS_persistent}). The no beam scenario was achieved by attenuating the 15 keV x-ray beam with 1-2 mm of aluminum, bringing the x-ray flux to effectively zero. The choice of fitting persistent current to decaying exponential curves was an ansatz attained from viewing plots of total current data. Using equation \ref{eq:persistent_fit}, exponential fits to the falling tails of both 400 V scans showed good agreement with data, suggesting that persistent current is likely proportional to the total current. An analytic continuation of the no beam fit was extrapolated back to the end of the full beam scan (Figure \ref{fig:DLS_persistent_all}). The intersection between the no beam exponential fit and full beam data, when the beam was attenuated to mostly zero flux in-between scans, supports the hypothesis that persistent current in the absence of photocurrent follows an exponential decay over time. A simple exponential may be an insufficient model for the non-central pad active area under beam, since the falling tail of the full beam scan is a combination of both time-dependent persistent current and position-dependent beam-induced signal. As a straightforward comparison between full and no beam results however, applying an expoential fit to both data sets is sufficient for comparison (Eq. \ref{eq:persistent_fit}, Table \ref{persistent_table}).

\begin{equation}
I_{total}(t)=Ae^{-kt}+C
\label{eq:persistent_fit}
\end{equation}

Persistent current is an additional background that was not accounted for during background subtraction method. Fully accounting for persistent current requires parameterizing exponential variables to bias voltage, beam flux, and other possible parameters. Recreating the same experimental conditions at DLS and focusing on investigating persistent current would resolve how prevalent this phenomena is in all other scans at DLS. 

While the impact on signal width is noticeable, the impact of persistent current on photocurrent uniformity at low voltages is relatively small. The value of the constant k implies only O(2\%) current change during the scan across the span of the central pad area (Table \ref{persistent_table}). Additionally, the current variation in the central pad region due to this phenomenon can be approximated as a linear displacement constrained to the central pad area, which is already accounted for in background subtraction. The impact of persistent current on photocurrent behavior measurements then is minimal. Determining if persistent current is an inherent property of InP and if persistent current would affect InP detector performance in high-occupancy environments are subjects left to future studies.

\begin{table}
\centering
\caption{Persistent Current Fit Parameters at 400 V}
\begin{tabular}{lrrr}
\hline
Beam State & A [nA/ai1] & k [$s^{-1}$] & $C$ [nA/ai1] \tabularnewline
\hline
Full & 845 & .00652 & 3370\tabularnewline
Attenuated & 113 & .00113 & 2670\tabularnewline
\hline
\end{tabular}
\label{persistent_table}
\end{table}

\subsubsection{High Voltage Instability}
\label{subsubsectionDLSHV}

\begin{figure}
    \begin{subfigure}[b]{0.5\textwidth}
        \centering
        \includegraphics[width=\textwidth]{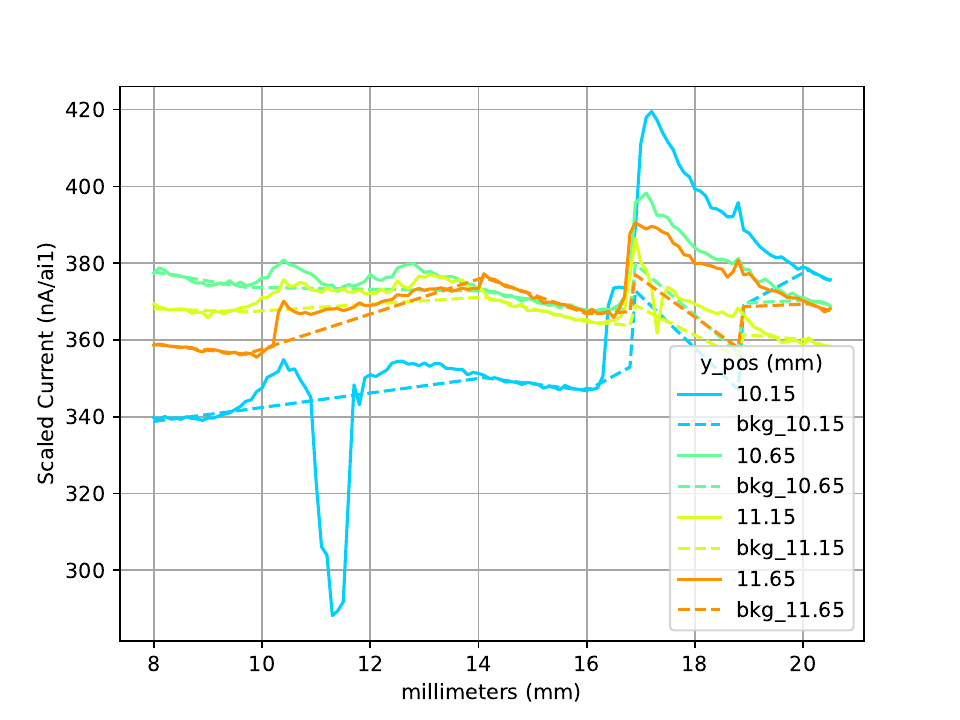}
        \caption{60 V}
        \label{fig:DLS_HV_60}
    \end{subfigure}
    \begin{subfigure}[b]{0.5\textwidth}
        \centering
        \includegraphics[width=\textwidth]{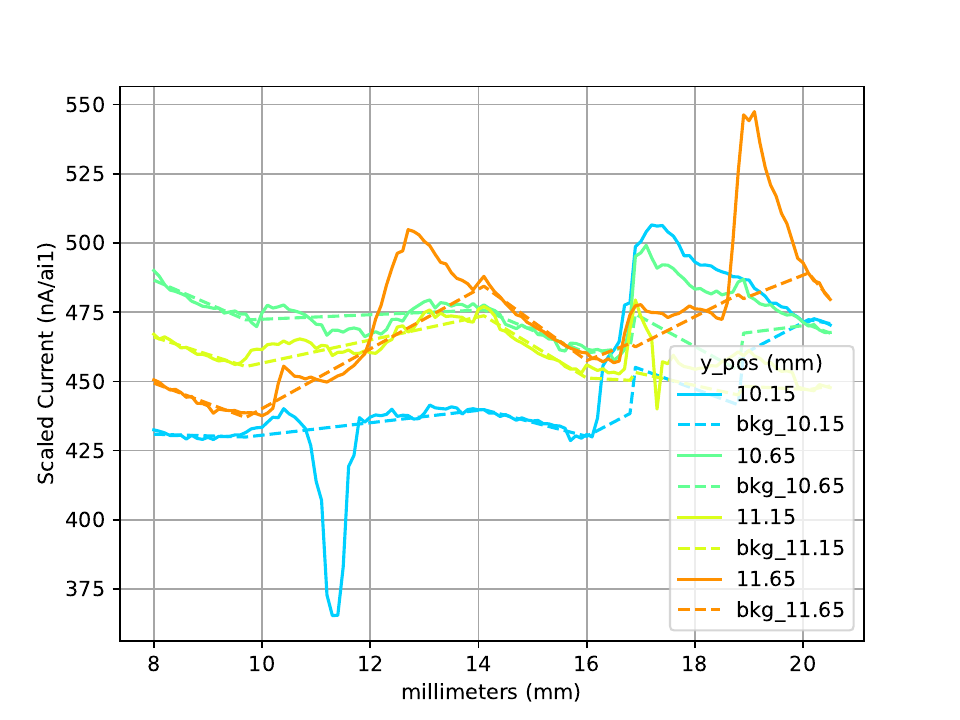}
        \caption{70 V}
        \label{fig:DLS_HV_70}
    \end{subfigure}
    
    \caption{Total Current and Background Fits for High Voltage Line Scans. Device 1 and 2 are located around 11 mm and 18 mm respectively on the x-axis}
    \label{fig:DLS_HV}
\end{figure}

Line scans taken at 60 and 70 V, defined as the high voltage line scans, show aberrant behavior that makes photocurrent assessment difficult (Figure \ref{fig:DLS_HV}). On Device 1, the photocurrents are much lower and more diffuse in position than those seen at lower voltages. By contrast, Device 2 suffers from an orthogonal issue of a prevalent exponentially decaying current, potentially due to persistent current, likely unrelated to photocurrent. While the cause of both of these phenomena is not known, their presence motivates removing 60 and 70 V line scans from active area assessment.

\subsubsection{Accounting for Background Noise}
\label{subsubsectionDLSreadout}

\begin{figure}
    \begin{subfigure}[b]{0.5\textwidth}
        \centering
        \includegraphics[width=\textwidth]{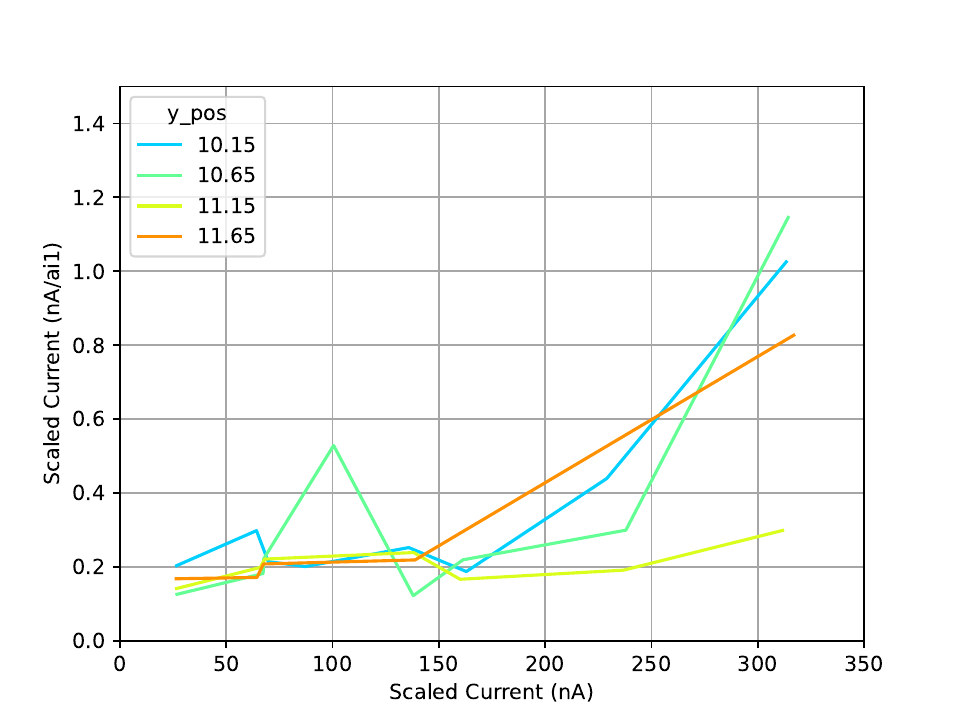}
        \caption{Background}
        \label{fig:DLS_std_0}
    \end{subfigure}
    \begin{subfigure}[b]{0.5\textwidth}
        \centering
        \includegraphics[width=\textwidth]{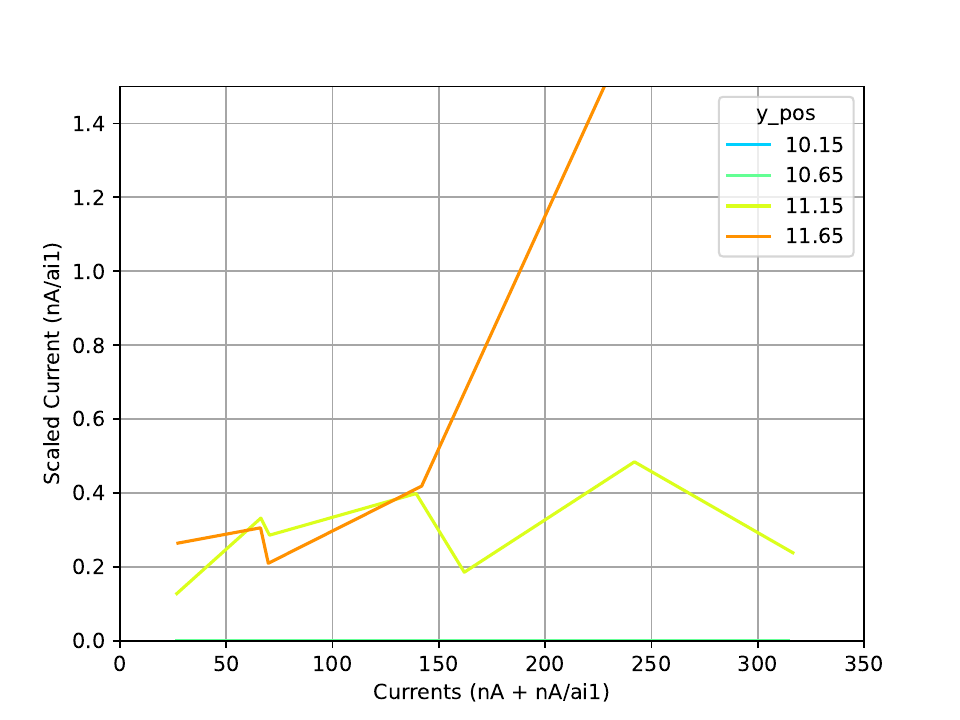}
        \caption{Device 1}
        \label{fig:DLS_std_1}
    \end{subfigure}
    \begin{subfigure}[b]{0.5\textwidth}
        \centering
        \includegraphics[width=\textwidth]{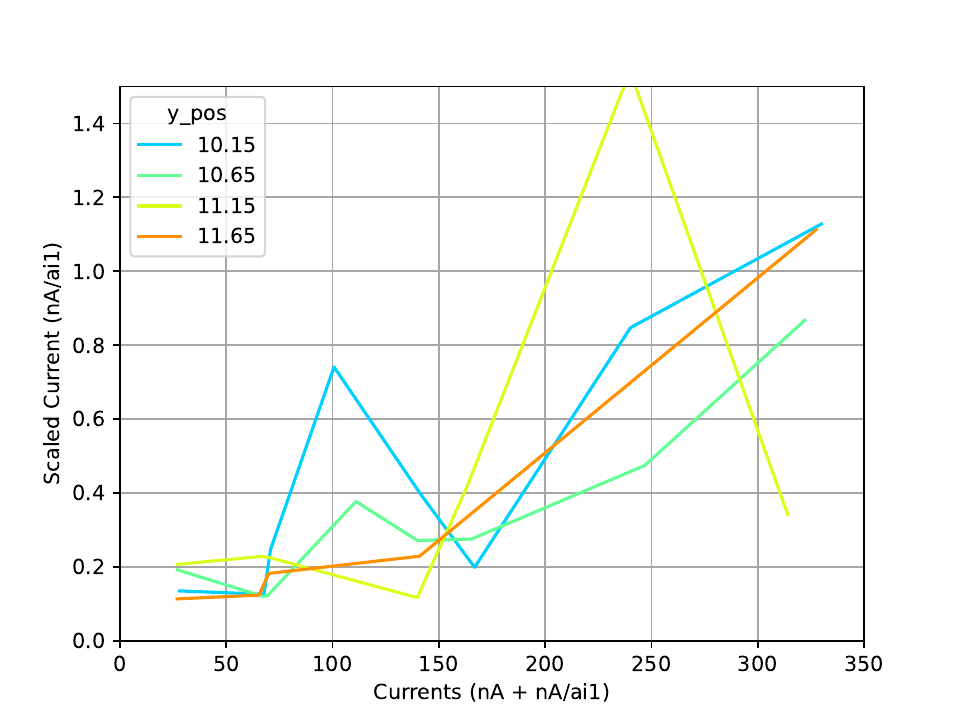}
        \caption{Device 2}
        \label{fig:DLS_std_2}
    \end{subfigure}
    \begin{subfigure}[b]{0.5\textwidth}
        \centering
        \includegraphics[width=\textwidth]{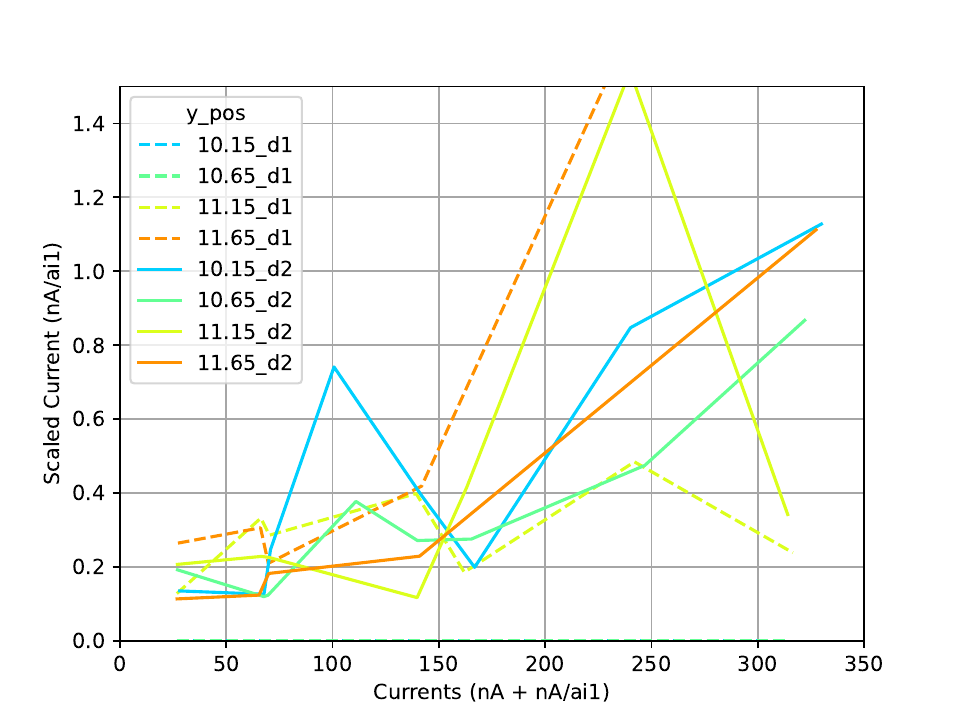}
        \caption{Device 1 and 2}
        \label{fig:DLS_std_all}
    \end{subfigure}
    
    \caption{Standard Deviation over Total Current is plotted in three regions: Background, Device 1, and Device 2. Background refers to the non-photocurrent region from x = 14.5 to 15.5 mm. Device 1 and Device 2 refer to the central pads of device 1 and 2, respectively. Below-center and center lines for Device 1 are not shown in Figure (b) due to the dip current affecting associated line scans. High variance was seen for Device 1 at the edge line likely due to edge effects at/near the edge of the active area (Figure \ref{fig:DLS_area20})}
    \label{fig:DLS_std}
\end{figure}

Evaluating active area uniformity requires accounting for non-photocurrent sources of noise as well. Comparing the standard deviation (SD) as a function of total current in the non-photocurrent region, device 1 central pad, and device 2 central pad shows similarity in both the magnitude and trend of SD (Figure \ref{fig:DLS_std}). The dominant source in non-uniformity seen in active areas then is likely due to readout noise, whether that be from thermal noise, an adaptive measurement range in the power supply, or other sources otherwise unrelated to photocurrent. In some cases, the standard deviation found in central pad regions were lower than that found in background. Subtracting background noise from central pad noise in quadrature then may not be possible for some data points. While RSD (relative standard deviation) will be reported in for DLS results, conclusions on InP response uniformity from them may prove difficult. At minimum, it can be safely reasoned that variance in InP uniformity contributes at the same order as variance from orthogonal sources.

\subsubsection{Results}
\label{subsubsectionDLSLine}

\begin{figure}
    \begin{subfigure}[b]{0.5\textwidth}
        \centering
        \includegraphics[width=\textwidth]{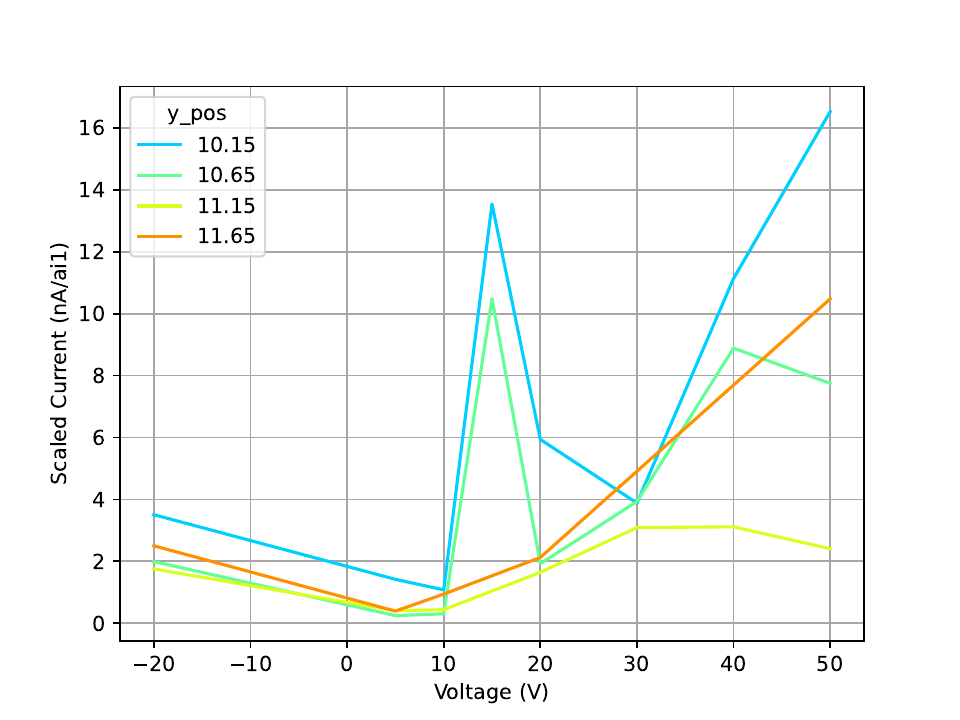}
        \caption{Photocurrent Magnitude vs Voltage}
        \label{fig:DLS_magnitude2}
    \end{subfigure}
    \begin{subfigure}[b]{0.5\textwidth}
        \centering
        \includegraphics[width=\textwidth]{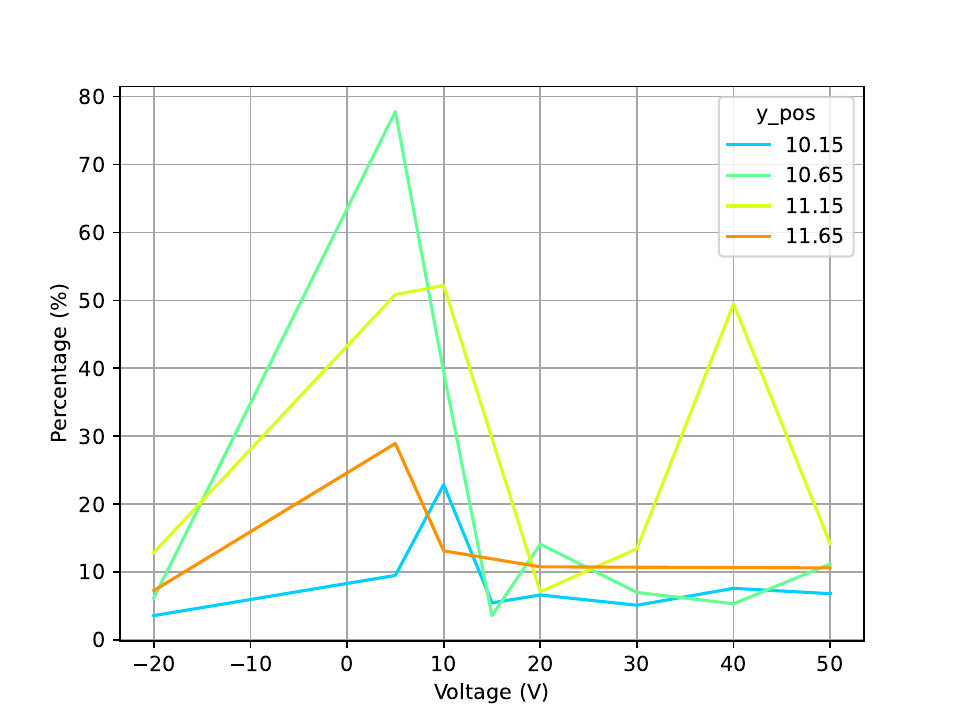}
        \caption{RSD vs Voltage}
        \label{fig:DLS_rsd2}
    \end{subfigure}
    \begin{subfigure}[b]{0.5\textwidth}
        \centering
        \includegraphics[width=\textwidth]{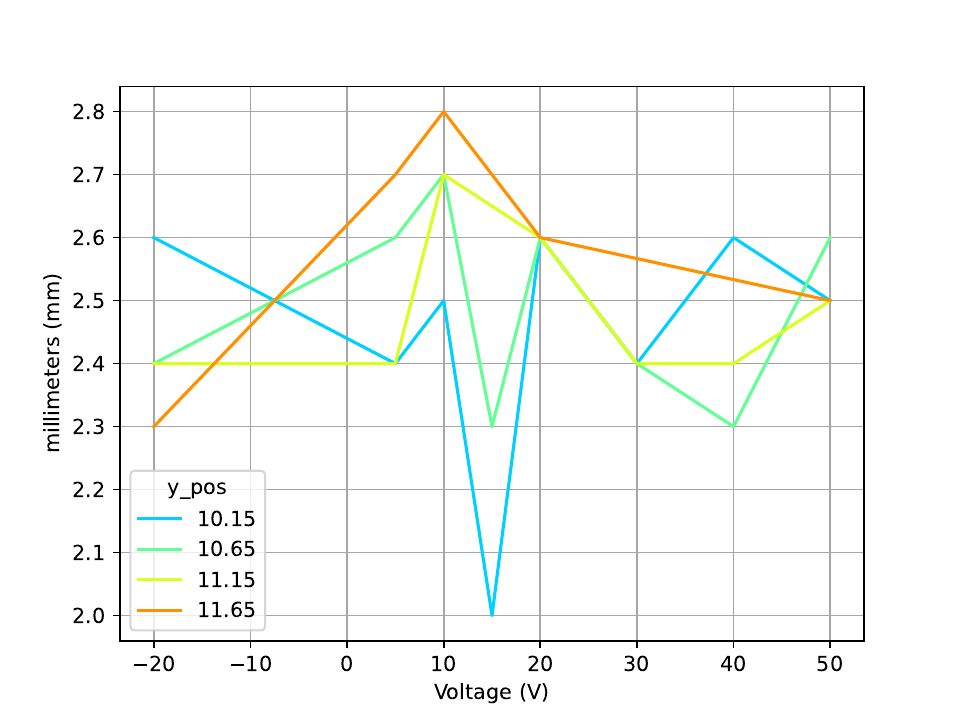}
        \caption{Width vs Voltage}
        \label{fig:DLS_width2}
    \end{subfigure}
    \begin{subfigure}[b]{0.4\textwidth}
        \centering
        \includegraphics[width=\textwidth]{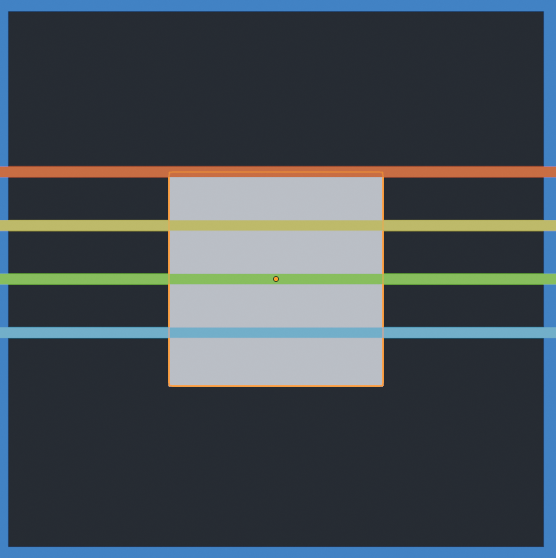}
        \caption{Color-mapping to Y-position}
        \label{fig:DLS_colorcode}
    \end{subfigure}
    
    \caption{Figures (a-c) detail properties of Device 2 active area photocurrent. Relative standard deviation (RSD) is defined in the central pad area as the photocurrent standard deviation divided by the photocurrent magnitude, giving the active area non-uniformity as a percentage of the active area magnitude. \\In Figure (d), the below-center, center, above-center, and edge lines were positioned to cut through both devices' active areas. The device is represented as a black square with a white central pad, with the guard ring omitted for visual clarity.}
    \label{fig:DLS_stats}
\end{figure}

Only horizontal scans which swept over the central pad area, corresponding to four stage y-positions y = 10.15, 10.65, 11.15, and 11.65 mm (below-center, center, above-center, edge), are used for evaluating photocurrent uniformity, width, and magnitude (Figure \ref{fig:DLS_stats}). Due to issues found in Device 1 (dip current at lower y-position scans and higher y-position scans lying at the active area edge), only Device 2 will be evaluated. While included in plots, data at 5 V is imprecise due to low SN ratio between photocurrent and ambient noise. Results listed were obtained after applying the same background subtraction method detailed in \ref{subsubsectionCLSLine}.

The larger voltage range at DLS, compared to CLS, alleviates previous concerns about extrapolating trends in this data set. Photocurrent magnitude tends to rise as a function of bias voltage, while RSD and width remain near constant. Photocurrent magnitude looks to have a minima at the center of the central pad, rising until scans reach the central pad edge. This is corroborated in both Figure \ref{fig:DLS_magnitude2} and horizontal view of scans at Figure \ref{fig:DLS_iy_10-15}. RSD and width also remain near constant across scan position. Evaluating RSD is difficult due to background noise, though the trend seen in Figure \ref{fig:DLS_rsd2} implies that device response variation does not exceed the readout noise. The implied shape of the active area is of a plateau that is nearly flat across position, rising in magnitude further from the center due to edge effects. As a function of bias voltage, this active area rises in magnitude but maintains the same shape.

Parameters at both 5 V, 15 V, and the above-center line are worth noting for not following nominal trends in photocurrent parameters. The low-voltage (up to 5 V) SN ratio at DLS is worse than at CLS despite smaller noise due to a lower photocurrent, stemming from lower temperatures during operation (Table \ref{scan_param_table}). Alongside differences in beam width and beam current between experiments, a direct one-to-one comparison of uniformity between either facility may not be valid. As for 15 V, since the plateau seen at both scan y-positions is clear, without noise or drastic discontinuities from background subtraction, causes for the small width and RSD yet high photocurrent remain unclear. The data collected at the above-center line, corresponding to yellow data in Figure \ref{fig:DLS_stats}, exhibits high RSD due to an anomalous spike in current at x = 17.1 mm which persists across scan days. Unlike dip current, this spike only appears at voltages greater than or equal to 30 V with consistent amplitude.

Temperature variation at DLS was no more than 1°C on each day. Therefore, temperature fluctuations are unlikely to significantly contribute to any phenomena mentioned (noise, persistent current, high voltage instability).

\subsection{Photocurrent Comparison}

A brief comparison of the signal magnitude between CLS and DLS is performed, using known beam flux values and photocurrent measurements from InP. Large uncertainties and a lack of focus on studying this comparison in both beam tests makes a precision comparison between results from these facilities impossible with existing data.

From Table \ref{beam_table}, the ratio in beam flux between CLS and DLS is calculated to be 1060. A similar ratio can be made between the photocurrents at both facilities to check on the signal scaling with the flux. In the absence of the charge trapping inside the device, the observed photocurrent is directly proportional to beam flux by the below equation:

\begin{equation}
I = \Phi_{\gamma}P_{\mathrm{absorption}}A_{\mathrm{signal}}
\label{test_equation}
\end{equation}

Here, $\Phi_{\gamma}$ corresponds to beam flux, $P_{\mathrm{absorption}}$ is the probability of absorption into the material, and $A_{\mathrm{signal}}$ is the ionization signal. Ionization signal is approximately equal to the ratio between the photon energy $E_{\mathrm{beam}}$ and the electron/hole pair production energy: 

\begin{equation}
A_{\mathrm{signal}} ~= E_{\mathrm{beam}} / E_{\mathrm{e-h}}
\label{ionization_signal}
\end{equation}

Since the main difference in equation \ref{test_equation} between both facilities is the beam flux, the photocurrent measurement can serve as a proxy for the beam flux. The additional operational temperature difference of 17.5 C seen in Table \ref{beam_table} will have a negligible effect on the absorption probability.

As shown in Figure \ref{fig:DLS_magnitude2}, the photocurrent across all DLS line scans is approximately 1 nA at 10 V. This number was scaled down by the flux factor of 1.18 during the data analysis. Similarly, in Figure \ref{fig:CLS_photocurrent}, the average CLS photocurrent magnitude (scaled down by the flux factor of 1.25) between Device 1 and Device 2 is approximately 0.07 $\mu$A at 2 V. Assuming a linear relationship between current and voltage at low applied voltage magnitude, the photocurrent at 10 V can be extrapolated to be 0.35 $\mu$A. After accounting for the flux factors, the inherent ratio of the current signals is estimated to be 375.

We consider the ratio of the fluxes and the ratio of derived signals to be fairly similar, given the 3 orders of magnitude difference and a number of uncertainties involved in the derivations. The measurements themselves have the uncertainties. The photon fluxes reported were measured with different methods, the beam optics in one case and a calibration diode in another. The linear extrapolation of CLS photocurrent is imperfect, noting the non-linear photocurrent behavior over voltage observed in Figure \ref{fig:DLS_magnitude2}. Given that the non-linear behavior of observed photocurrent at DLS points to a true 10 V CLS photocurrent being greater than the linear estimate, the measurement improvement may bring the ratios closer together.

At future beam tests, we could improve this comparison by addressing the high fluctuation issues seen at high voltages discussed in Section \ref{subsubsectionCLSLineResults} and measuring the photocurrents at the same voltages. The photon flux ratio comparison could be checked by self-calibrating the flux using the same Si diode at both facilities, eliminating uncertainties associated with their estimates at the facilities.

The equations~\ref{test_equation} and~\ref{ionization_signal} allow a direct estimate of the currents expected. The values are much higher than the measurements. For example, for the DLS flux value we estimate the signal of 67 nA. This difference with the measurements and the strong dependence of the signal on the high voltage value indicate that the signal trapping takes place during the charge transport to the readout electrodes. We will explore this topic in a forthcoming publication with a direct measurement of ionizing signal from a beta source.

\section{Summary and Conclusions}

Novel ionization-sensitive devices based on InP material have been fabricated. They conform with the expected planar and stackup geometries. We performed several tests of their electrical characteristics as a measure of material properties.
InP:Fe devices typically have a linear IV dependence up to $\pm$ 350 V, which is symmetric with regard to the voltage polarity. Out of a total 48 devices measured, a cluster of 3 devices neighboring on the original wafer showed up to 20\% higher leakage currents while maintaining nominal capacitances. The relative current variation between the rest of the devices is only 9\%. A device taken to breakdown reached it at 1kV, which corresponds to 28.9 kV/cm field strength. 

The measured capacitance is somewhat larger than the expectations based on parallel plate approximation with equal plates, likely due to the large metallized area on the back. We observe the bias voltage dependence for positive polarity and low frequency, which we attribute to the presence of the surface charges on the top-side un-metallized area.
While normally exhibiting a capacitance of 2 pF, the capacitance can be further dropped and edge effects minimized with grounded guard rings. The charged particle detection time window should ideally take place on the order of microseconds or shorter, as capacitances recorded at frequencies lower than 1 MHz showed voltage dependence, both in magnitude and polarity.

We additionally tested the devices in micro-focused X-ray beams at CLS and Diamond facilities under sub-zero temperatures. Response variation in the active area under the central pad is approximately 10\% within one relative standard deviation, with the most likely dominant source coming from readout noise. The response magnitude is a strong function of the applied voltage, indicating a reduction of the effective charge trapping for higher fields due to faster drift.

Edge effects affected the response much more than noise, with higher signals found closer to the central pad edges. Adding more guard rings to the layout and using grounded guard rings in future beam experiments could demonstrate how much this source of non-uniformity could be mitigated. A thicker top metal may attenuate this phenomenon by reducing peak electric field at the edges. 

Regions of high current were identified, such as the dip current on Device 1 and spike at the above-center line on Device 2. The leading hypothesis for their cause is damage from wirebonding, as additional force was required to attach wires to the thin central pad metal layer. If this hypothesis holds true, future device architecture could account for this fault with thicker metallization. 

Observations made at both CLS and DLS could only probe the upper layers of the device with a beam energy of 15 keV. Testing device responses at higher beam energies could allow for full device characterization (Figure \ref{fig:Xray_absorption}). 

The difference of the measurements with the expected values from the photon fluxes and the strong dependence of the signal on the high voltage value indicate a presence of charge trapping in the devices. We will explore this topic in a forthcoming publication with a direct measurements of the ionization from beta particles.

As one of the first evaluations of InP:Fe properties for charged particle detection applications, these experiments provided a picture of a material with high breakdown voltage and good uniformity within the available statistics: between the devices in leakage current and capacitance, and within devices in response to ionizing radiation. The samples used featured a vulnerability to edge effects which can be addressed in subsequent fabrications.

\acknowledgments
This project was partially funded by a seed grant from the Office of Research at University of California, Santa Cruz. J. Ott would like to thank the Finnish Cultural Foundation for financial support through the Postdoc Pool of Finnish Foundations. 
This work was also supported by the Diamond Light Source (proposal OM32397) and the Canadian Light Source (proposals 36G12506 and 38G13372). In addition, the work was supported by the Canada Foundation for Innovation (CFI) and the Natural Science and Engineering Research Council of Canada (NSERC).
Funding from Argonne National Laboratory, provided by the Director, Office of Science, of the U.S. Department of Energy under Contract No. DE-AC02-06CH11357. This work was supported by the Office of Naval Research (Contract: N000141812583) and University of Illinois at Chicago. Work performed at the Center for Nanoscale Materials, U.S. Department of Energy Office of Science User Facilities, was supported by the U.S. DOE, Office of Basic Energy Sciences, under Contract No. DE-AC02-06CH11357.

\bibliography{InP_Canada_tb_bibfile}{}
\bibliographystyle{JHEP}

\end{document}